\documentclass[dvipsnames]{aastex62}

\usepackage{amsmath}
\usepackage[normalem]{ulem}

\newcommand{\y}{\color{blue}}

\graphicspath{{./}{figures/}}

\received{}
\revised{}
\accepted{}
\begin{document}

\title{The surprisingly low carbon mass in the debris disk around HD\,32297}

\correspondingauthor{Gianni Cataldi}
\email{cataldi.gia@gmail.com}

\author[0000-0002-2700-9676]{Gianni Cataldi}
\altaffiliation{International Research Fellow of Japan Society for the Promotion of Science (Postdoctoral Fellowships for Research in Japan (Standard)).}
\affil{Department of Astronomy \& Astrophysics, University of Toronto, 50 St.\ George Street, Toronto, ON M5S 3H4, Canada}
\affil{Subaru Telescope, National Astronomical Observatory of Japan, 650 North Aohoku Place, Hilo, HI 96720, USA}
\affil{National Astronomical Observatory of Japan, Osawa 2-21-1, Mitaka, Tokyo 181-8588, Japan}
\affil{Department of Astronomy, Graduate School of Science, The University of Tokyo, 7-3-1 Hongo, Bunkyo-ku, Tokyo 113-0033, Japan}
\affil{Konkoly Observatory, Research Centre for Astronomy and Earth Sciences, Konkoly-Thege Mikl\'os \'ut 15--17, 1121 Budapest, Hungary}

\author{Yanqin Wu}
\affil{Department of Astronomy \& Astrophysics, University of Toronto, 50 St.\ George Street, Toronto, ON M5S 3H4, Canada}

\author{Alexis Brandeker}
\affil{Department of Astronomy, Stockholm University, AlbaNova University Center, 106 91, Stockholm, Sweden}

\author{Nagayoshi Ohashi}
\affil{Subaru Telescope, National Astronomical Observatory of Japan, 650 North Aohoku Place, Hilo, HI 96720, USA}

\author{Attila Mo{\'o}r}
\affil{Konkoly Observatory, Research Centre for Astronomy and Earth Sciences, Konkoly-Thege Mikl\'os \'ut 15--17, 1121 Budapest, Hungary}
\affil{ELTE E\"otv\"os Lor\'and University, Institute of Physics, P\'azm\'any P\'eter s\'et\'any 1/A, 1117 Budapest, Hungary}

\author{G\"oran Olofsson}
\affil{Department of Astronomy, Stockholm University, AlbaNova University Center, 106 91, Stockholm, Sweden}

\author{P\'eter \'Abrah\'am}
\affil{Konkoly Observatory, Research Centre for Astronomy and Earth Sciences, Konkoly-Thege Mikl\'os \'ut 15--17, 1121 Budapest, Hungary}
\affil{ELTE E\"otv\"os Lor\'and University, Institute of Physics, P\'azm\'any P\'eter s\'et\'any 1/A, 1117 Budapest, Hungary}

\author{Ruben Asensio-Torres}
\affil{Department of Astronomy, Stockholm University, AlbaNova University Center, 106 91, Stockholm, Sweden}

\author{Maria Cavallius}
\affil{Department of Astronomy, Stockholm University, AlbaNova University Center, 106 91, Stockholm, Sweden}

\author{William R.F. Dent}
\affil{ALMA Santiago Central Offices, Alonso de Cordova 3107, Vitacura, Santiago, Chile}

\author{Carol Grady}
\affil{Eureka Scientific, 2452 Delmer, Suite 100, Oakland CA 94602-3017, USA}

\author{Thomas Henning}
\affil{Max Planck Institute for Astronomy, K\"onigstuhl 17, 69117 Heidelberg, Germany}

\author{Aya E.\ Higuchi}
\affil{National Astronomical Observatory of Japan, Osawa 2-21-1, Mitaka, Tokyo 181-8588, Japan}

\author{A.\ Meredith Hughes}
\affil{Astronomy Department and Van Vleck Observatory, Wesleyan University, 96 Foss Hill Drive, Middletown, CT 06459, USA}

\author{Markus Janson}
\affil{Department of Astronomy, Stockholm University, AlbaNova University Center, 106 91, Stockholm, Sweden}

\author{Inga Kamp}
\affil{Kapteyn Astronomical Institute, University of Groningen, P.O.\ Box 800, 9700 AV Groningen, The Netherlands}

\author{\'Agnes K\'osp\'al}
\affil{Konkoly Observatory, Research Centre for Astronomy and Earth Sciences, Konkoly-Thege Mikl\'os \'ut 15--17, 1121 Budapest, Hungary}
\affil{ELTE E\"otv\"os Lor\'and University, Institute of Physics, P\'azm\'any P\'eter s\'et\'any 1/A, 1117 Budapest, Hungary}
\affil{Max Planck Institute for Astronomy, K\"onigstuhl 17, 69117 Heidelberg, Germany}

\author{Seth Redfield}
\affil{Astronomy Department and Van Vleck Observatory, Wesleyan University, 96 Foss Hill Drive, Middletown, CT 06459, USA}

\author{Aki Roberge}
\affil{Exoplanets and Stellar Astrophysics Laboratory, NASA Goddard Space Flight Center, Greenbelt, MD 20771, USA}

\author{Alycia Weinberger}
\affil{Department of Terrestrial Magnetism, Carnegie Institution of Washington, 5241 Broad Branch Road NW, Washington, DC 20015-1305, USA}

\author{Barry Welsh}
\affil{Space Sciences Laboratory, UC Berkeley, 7 Gauss Way, Berkeley, CA 94720, USA}




\begin{abstract}
Gas has been detected in a number of debris disks. It is likely secondary, i.e.\ produced by colliding solids. Here, we report ALMA Band 8 observations of neutral carbon in the CO-rich debris disk around the 15--30\,Myr old A-type star HD\,32297. We find that C$^0$ is located in a ring at $\sim$110\,au with a FWHM of $\sim$80\,au, and has a mass of $(3.5\pm0.2)\times10^{-3}$\,M$_\oplus$. Naively, such a surprisingly small mass can be accumulated from CO photo-dissociation in a time as short as $\sim$10$^4$\,yr. We develop a simple model for gas production and destruction in this system, properly accounting for CO self-shielding and shielding by neutral carbon, and introducing a removal mechanism for carbon gas. We find that the most likely scenario to explain both C$^0$ and CO observations, is one where the carbon gas is rapidly removed on a timescale of order a thousand years and the system maintains a very high CO production rate of $\sim$15\,M$_\oplus$\,Myr$^{-1}$, much higher than the rate of dust grind-down. We propose a possible scenario to meet these peculiar conditions: the capture of carbon onto dust grains, followed by rapid CO re-formation and re-release. In steady state, CO would continuously be recycled, producing a CO-rich gas ring that shows no appreciable spreading over time. This picture might be extended to explain other gas-rich debris disks.
\end{abstract}


\section{Introduction} \label{sec:intro}
A considerable fraction of main-sequence stars are surrounded by dusty disks known as \emph{debris disks} \citep[e.g.][]{Hughes_etal_2018}. They are analogous to the asteroid belt and the Kuiper belt in the solar system, and are often interpreted as leftover products from the planet formation era. The observed dust is secondary, i.e.\ produced from the collisional destruction of larger (asteroidal or cometary) bodies. Observations of debris disks can constrain the composition of these bodies as well as the evolution and dynamics of extrasolar systems.

A sub-sample of debris disks, mainly surrounding young A-type stars, are observed to contain gas besides the dust. The Atacama Large Millimeter/submillimeter Array (ALMA) has been instrumental in increasing the sample of gaseous debris disks by providing sensitive observations of CO emission \citep[e.g.][]{Kospal_etal_2013,Lieman-Sifry_etal_2016,Moor_etal_2017}. Today, about 20 gaseous debris disks are known. The observed CO masses show a large spread. In some systems, the CO column density is so low that CO is not self-shielded against photodissociation by the interstellar radiation field (ISRF) \citep[e.g.][]{Marino_etal_2016,Matra_etal_2017_Fomalhaut,Matra_etal_2019}. As a consequence, CO is photodissociated on a short timescale \citep[$\sim$120\,yr,][]{Visser_etal_2009} and is therefore secondary, produced from the destruction of icy, cometary bodies \citep[e.g.][]{Kral_etal_2017}. On the other hand, a few disks contain CO masses comparable to protoplanetary disks \citep[e.g.][]{Kospal_etal_2013,Moor_etal_2017}. Such high CO masses have been difficult to explain in a secondary scenario. Thus, it was suggested that these are 'hybrid' disks, i.e.\ disks where secondary dust and primordial (i.e.\ leftover from the protoplanetary phase) gas co-exist \citep{Kospal_etal_2013,Moor_etal_2017,Pericaud_etal_2017}. However, \citet{Kral_etal_2019,Marino_etal_2020} recently suggested that even these disks can be explained by secondary gas production if shielding of CO by neutral carbon is taken into account. Carbon is continuously produced from CO photodissociation. Eventually, the carbon column density can be high enough to attenuate dissociation of CO, allowing a large CO mass to build up.

This process might be at work in the debris disk around HD\,32297. Located at a distance of $132.3\pm1.0$\,pc \citep{BailerJones_etal_2018}, its spectral type has been described as A5--A7\footnote{\citet{Torres_etal_2006} determined the spectral type to be A0, but this was found too hot in several subsequent papers.} \citep{Fitzgerald_etal_2007,Donaldson_etal_2013,Rodigas_etal_2014} and its age has been estimated to be $\gtrsim15$\,Myr \citep{Rodigas_etal_2014} and $<30$\,Myr \citep{Kalas_etal_2005}. The dust component of the disk has been studied in considerable detail, both in scattered light \citep[e.g.][]{Boccaletti_etal_2012,Currie_etal_2012,Asensio-Torres_etal_2016} and thermal emission \citep[e.g.][]{Maness_etal_2008,Donaldson_etal_2013,MacGregor_etal_2018}. Gas has been observed in the form of CO \citep{Greaves_etal_2016,MacGregor_etal_2018}, \ion{C}{2} \citep{Donaldson_etal_2013} and \ion{Na}{1} \citep{Redfield_etal_2007}. Since the $^{12}$CO emission is optically thick, \citet{Moor_etal_2019} recently used observations of the $^{13}$CO and C$^{18}$O isotopologues to constrain the total CO mass. They argue that the large CO mass can indeed be explained by secondary CO production combined with C shielding.

As carbon is the direct descendent of CO, and because of the pivotal role it plays in shielding CO, observations of carbon are an important tool to improve our understanding of gaseous debris disks. In this paper, we present ALMA observations of neutral carbon emission towards HD\,32297. We attempt to explain the observations with a secondary gas production scenario. Our paper is structured as follows: in section \ref{sec:obs}, we describe our observations. In section \ref{sec:results_analysis}, we present the data analysis and determine the spatial distribution and mass of the neutral carbon. In section \ref{sec:gas_production_model}, we present a model of the gas production and destruction. We discuss our results in section \ref{sec:discussion} and summarize in section \ref{sec:summary}.

\section{Observations and data reduction}\label{sec:obs}
The HD\,32297 disk was observed in a single pointing with ALMA Band 8 receivers on May 22, 2018, during ALMA Cycle 5 (project ID 2017.1.00201.S, PI: Cataldi). An array of 47 antennas arranged in a compact configuration was employed, with baselines ranging from 15 to 314\,m. The total integration time was 45\,min (with 18\,min on HD\,32297) and the median precipitable water vapor (pwv) was 0.6\,mm.

The spectral setup consisted of three spectral windows centered at 480, 482 and 494\,GHz, each with a bandwidth of 2\,GHz and 128 channels to observe the dust continuum. The fourth spectral window was covering the \ion{C}{1}~$^3$P$_1$--$^3$P$_0$ line at 492.16\,GHz, with 1920 channels, a channel spacing of 488\,kHz (0.30\,km\,s$^{-1}$) and an effective spectral resolution\footnote{see ALMA Cycle 6 Technical Handbook, Table~5.2, \url{https://almascience.nrao.edu/documents-and-tools/cycle6/alma-technical-handbook}} of 0.34\,km\,s$^{-1}$ (spectral averaging factor $N=2$).

The data were calibrated by the ALMA Science Pipeline within the Common Astronomy Software Applications package (CASA) 5.1.1 \citep{McMullin_etal_2007}. J0522-3627 (bandpass, flux) and J0505+0459 (phase) were observed as calibration sources.

\begin{figure}
\plottwo{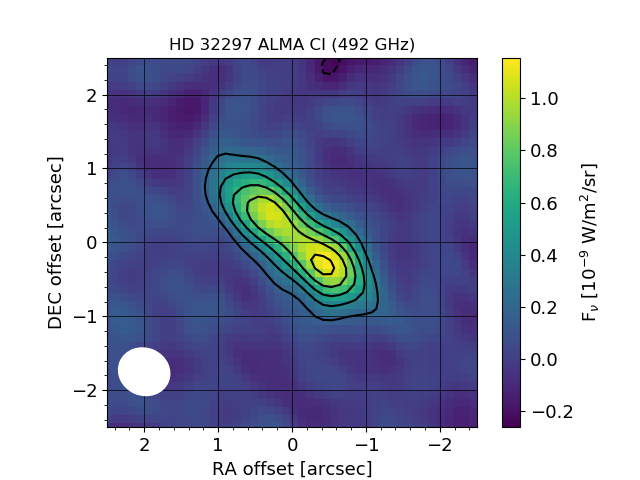}{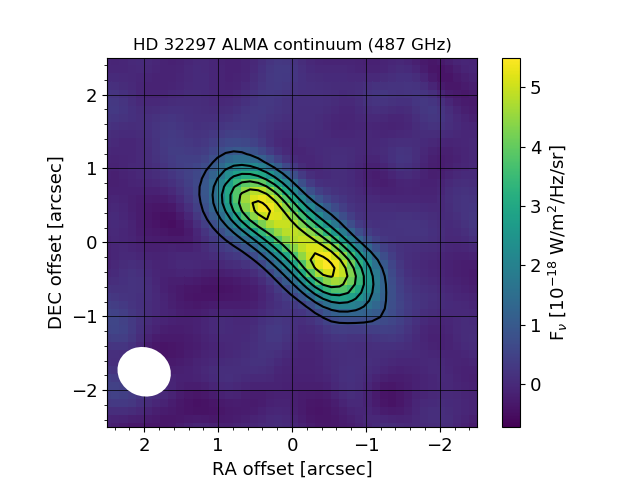}
\caption{Left panel: Moment 0 map of the \ion{C}{1} 492\,GHz emission observed towards HD\,32297. The stellar position is at $(0,0)$. Contours are drawn at intervals of 3$\sigma$, with $\sigma=7.3\times10^{-11}$\,W\,m$^{-2}$\,sr$^{-1}$. Right panel: ALMA Band 8 continuum image at 487\,GHz (616\,$\mu$m). Contours are drawn at intervals of 5$\sigma$, with $\sigma=1.7\times10^{-19}$\,W\,m$^{-2}$\,Hz$^{-1}$\,sr$^{-1}$. The synthesized beams are shown as white ellipses in the lower left.\label{fig:CI_mom0_continuum}}
\end{figure}

For further processing of the calibrated visibilities, we used CASA 5.3.0. We used the task \texttt{uvcontsub} to subtract the continuum emission from the spectral window covering the \ion{C}{1} line (the line was masked for the continuum fit). We produced an image cube by employing the CLEAN algorithm with the task \texttt{tclean} and natural weighting. The synthesized beam is $0.72\arcsec\times0.65\arcsec$ (96\,au $\times$ 86\,au) with a position angle of $67\arcdeg$. The achieved noise level (measured by taking the standard deviation of a sample of data points in a region of the cube without line emission) is $1.7\times10^{-17}$\,W\,m$^{-2}$\,Hz$^{-1}$\,sr$^{-1}$ (21\,mJy\,beam$^{-1}$). Figure~\ref{fig:CI_mom0_continuum} (left) shows the moment 0 map of the \ion{C}{1} line, produced by integrating the data cube over the (barycentric) velocities of 15.5--25.5\,km\,s$^{-1}$.

We also produced a continuum map (Fig.~\ref{fig:CI_mom0_continuum}, right) by imaging all spectral windows combined (with the \ion{C}{1} line masked) using the \texttt{tclean} task with natural weighting. The synthesized beam is $0.73\arcsec\times0.66\arcsec$ (97\,au $\times$ 88\,au) with a position angle of $66\arcdeg$. The noise level (measured as above) is $2.0\times10^{-19}$\,W\,m$^{-2}$\,Hz$^{-1}$\,sr$^{-1}$ (0.25\,mJy\,beam$^{-1}$).

The primary beam\footnote{see ALMA Cycle 6 Technical Handbook, section 3.2} at 492\,GHz has a full width at half maximum (FWHM) of $\sim$12\arcsec. This is significantly larger than the extent of the disk, so we do not apply a primary beam correction. Given our shortest baseline, the maximum recoverable scale\footnote{see ALMA Cycle 6 Technical Handbook, section 3.6} of our observations is $\sim$5\arcsec, i.e.\ larger than the disk's extent. Thus, no significant amount of flux should have been filtered out by the interferometer.

\section{Data Analysis and Results}\label{sec:results_analysis}

\subsection{Total \ion{C}{1} and continuum emission}
No significant asymmetries are visible in either the line or the continuum image. Their appearance is consistent with a ring viewed edge-on. We measure the total \ion{C}{1} and continuum emission using the images in figure \ref{fig:CI_mom0_continuum} by integrating all emission above 3$\sigma$. To estimate the error, we collect flux samples by shifting the integration region to parts of the image where no emission is seen. We adopt the standard deviation of the flux samples as our estimate of the statistical error. An additional 10\% flux calibration error \citep{Fomalont_etal_2014} is added in quadrature, and this turns out to be the dominant error source. We find a total 492\,GHz \ion{C}{1} flux of $(4.0\pm0.4)\times10^{-20}$\,W\,m$^{-2}$ and a total 487\,GHz (616\,$\mu$m) continuum flux of $(2.2\pm0.2)\times10^{-28}$\,W\,m$^{-2}$\,Hz$^{-1}$ ($22.0\pm2$\,mJy). Including previous continuum measurements from \textit{Herschel}/SPIRE\footnote{European Space Agency, 2017, Herschel SPIRE Point Source Catalogue, Version 1.0. \url{https://doi.org/10.5270/esa-6gfkpzh}} at 350 and 500\,$\mu$m and ALMA at 1.3\,mm \citep{MacGregor_etal_2018}, we derive a sub-millimeter spectral index of 2.4. This value is comparable to the millimeter spectral indices determined for a sample of debris disks by \citet{MacGregor_etal_2016}.

\subsection{Position-velocity diagram}
\begin{figure}
\plotone{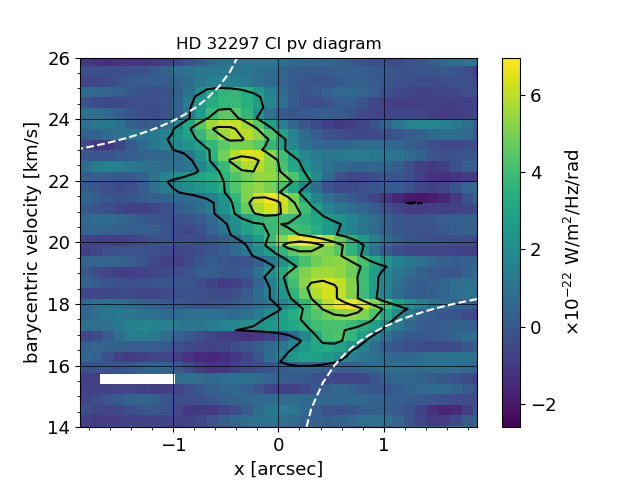}
\caption{Position-velocity diagram of the \ion{C}{1} emission. The $x$ coordinate runs along the major axis of the disk. Contours are drawn at intervals of 3$\sigma$. The spectro-spatial resolution is illustrated in the lower left by the white rectangle. The dashed curves show the tangential velocity for gas at projected distance $x$ from the star, assuming a stellar mass of 1.7\,M$_\odot$ and an inclination of 83.6\arcdeg.\label{fig:pv_diagram}}
\end{figure}

Figure~\ref{fig:pv_diagram} shows the position-velocity (pv) diagram of the \ion{C}{1} emission. We used the position angle measured in Appendix~\ref{appendix:PA} to rotate the data cube in order to align the midplane of the disk with the horizontal direction. The rotated data cube is then integrated from $-0.6\arcsec$ to $0.7\arcsec$ in the vertical direction $z$, where $z=0$ denotes the midplane\footnote{This slightly asymmetric integration is motivated by the vertical extent of the 3$\sigma$ contours in the moment 0 map.}. Analogous to the pv diagram for CO by \citet{MacGregor_etal_2018}, we also plot the Keplerian velocity curve, i.e.\ the extremum radial velocity that can be reached at projected distance $x$ from the star, assuming the same parameters as \citet{MacGregor_etal_2018}: a stellar mass of 1.7\,M$_\odot$ and an inclination of 83.6\arcdeg. In the absence of significant line broadening or resolution effects, one expects all emission to lie within these curves---as is observed. Hence, there is no evidence for sub-Keplerian rotation, contrary to the conclusion by \citet{MacGregor_etal_2018}.

\subsection{Modeling of the \ion{C}{1} emission}\label{sec:CI_modeling}
We proceed to constrain the spatial distribution and mass of neutral carbon by fitting models to the \ion{C}{1} data cube. For a given model, specified by the spatial distribution of the gas and the temperature profile, we first calculate the level population and gas emission in each cell of a three-dimensional grid by assuming local thermodynamic equilibrium (LTE). We verified that LTE is a reasonable approximation by calculating the full radiative transfer in non-LTE using the \texttt{LIME} code \citep{Brinch_Hogerheijde_2010} for a model that fits the data (Fig.~\ref{fig:highest_prop_Gaussian_ring}, see below). For that model, the total flux calculated in LTE was only a factor 1.3 higher than the flux from \texttt{LIME}. We use atomic data from the Leiden Atomic and Molecular Database \citep[LAMDA,][]{Schoier_etal_2005}. The line emission profile is assumed to be Gaussian with a broadening parameter of 1\,km\,s$^{-1}$. The emission is red- or blue-shifted according to the radial velocity (due to Keplerian rotation) of each grid cell. Then, the model is ray-traced as described in \citet{Cataldi_etal_2014} to account for optical depth. The resulting model cube is convolved in the spatial and spectral dimension to match the resolution of the data, and finally multiplied by the primary beam.

Short of a detailed heating-cooling analysis, we simply assume that the gas kinetic temperature behaves similarly as the local blackbody and varies with radial distance $r$ as $T(r) = T_{100}\sqrt{(100\,\mathrm{au})/r}$, where $T_{100}$ is a normalization constant obtained from the fit.  The gas vertical scale height is given by \citep[e.g.][]{Armitage_2009}
\begin{equation}\label{eq:scaleheight}
    H(r) = \sqrt{\frac{kTr^3}{\mu m_pGM_*}}
\end{equation}
where $k$ is the Boltzmann constant, $\mu$ the mean molecular weight, $m_p$ the proton mass, $G$ the gravitational constant and $M_*$ the stellar mass. We assume $\mu=14$ \citep[gas mass dominated by carbon and oxygen,][]{Kral_etal_2017}, although $\mu$ might be higher since the CO mass in HD\,32297 is large. The assumptions on the temperature and molecular weight should introduce only minor errors into our results.

We follow \citet{Kral_etal_2019} and consider a Gaussian surface density profile for neutral C:
\begin{equation}
    \Sigma_\mathrm{C^0}(r) = \Sigma_0\exp\left(-\frac{(r-r_0)^2}{2\sigma_r^2}\right)
\end{equation}
where $\Sigma_0$ is the surface density at $r=r_0$ and $\sigma_r$ describes the radial width of the ring-like mass distribution. The number density of neutral carbon is then given by
\begin{equation}
    n_\mathrm{C^0}(r{\y ,z})=\frac{\Sigma_\mathrm{C^0}(r)}{\sqrt{2\pi}Hm_\mathrm{C}}\exp\left(-\frac{z^2}{2H^2}\right)
\end{equation}
where $z$ is the distance from the mid-plane and $m_\mathrm{C}$ the mass of a C atom.

\begin{figure}
\plotone{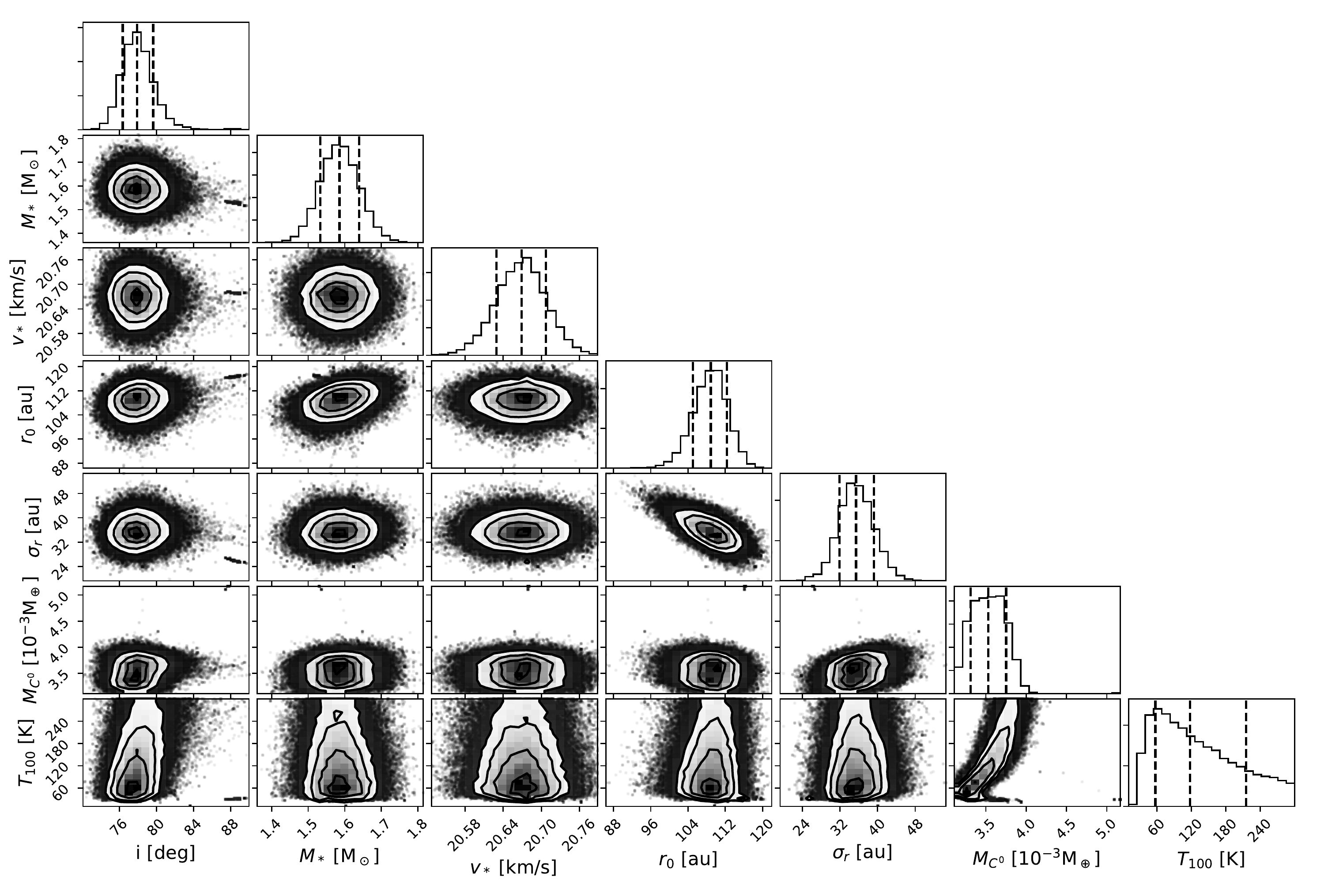}
\caption{Posterior distributions for selected parameters of the Gaussian ring model. The vertical dashed lines indicate the 16th, 50th and 84th percentile.\label{fig:cornerplot_Gaussian}}
\end{figure}

We use a Markov chain Monte Carlo (MCMC) method implemented by the \texttt{emcee} package \citep[version 2.2.1,][]{Foreman-Mackey_etal_2013} to fit the model to our data (in the image space). The correlated noise in the ALMA data is handled as described by \citet{Booth_etal_2017}. The following parameters are free to vary: the disk inclination $i$, the stellar mass $M_*$, the stellar radial velocity $v_*$, the ring location $r_0$, the ring width $\sigma_r$, the total mass of neutral carbon $M_\mathrm{C^0}$ and the temperature normalisation $T_{100}$. In addition, we also fit for astrometric offsets of the disk center, $\Delta x$ and $\Delta z$, which are parallel and perpendicular to the mid-plane respectively. For each parameter $\theta$, we assume uninformative priors, either with a location invariant density ($\pi(\theta)=\mathrm{const}$) or a scale invariant density ($\pi(\theta)\propto\theta^{-1}$), over the range given in Table~\ref{tab:fit_parameters}. The position angle is fixed to 48.5$\arcdeg$ as determined in Appendix~\ref{appendix:PA}. The MCMC is run for 1500 steps with 200 walkers. We discarded the first 200 samples of each walker. Table~\ref{tab:fit_parameters} shows derived parameters (50th percentile) together with error bars (16th and 84th percentile). Figure~\ref{fig:cornerplot_Gaussian} shows the posterior distribution of selected parameters. We note that the temperature is not well constrained by our data. This reflects the fact that the fractional population of the transition's upper level peaks at $\sim$30\,K and then stays approximately constant for higher temperatures. The relation between the mass and the temperature is governed by the level population, as expected for an optically thin gas. The derived inclination of $77.9\arcdeg^{+1.7}_{-1.5}$ is lower than previously derived values from dust observations with ALMA \citep[$83.6\arcdeg^{+4.6}_{-0.4}$,][]{MacGregor_etal_2018} or from scattered light \citep[$\sim$88\arcdeg,][]{Boccaletti_etal_2012,Currie_etal_2012}. The origin of this discrepancy is unclear, but one possibility is an asymmetry in the disk that our model does not account for. We run an additional MCMC simulation with a fixed inclination of 88\arcdeg and verified that the results for the other parameters do not change. In Fig.~\ref{fig:highest_prop_Gaussian_ring}, we present the fit with the highest posterior probability in our chain. No strong residuals are left, suggesting that our model is a good representation of the true gas distribution. This model has a peak optical depth of 0.2 along the line of sight. 

We note that the calibration uncertainty, not included in the MCMC fit, might introduce some additional uncertainty in the fitted parameters.

In summary, the C$^0$ gas distribution can be fit with a ring centered at $\sim$110\,au with a FWHM of $\sim$80\,au. In Appendix~\ref{appendix:power_law_models}, we also explore fitting the data using power-law models, to accommodate the possibility that the gas has spread into a broad disk. Our results there confirm that the gas is indeed distributed in a ring-like geometry. In Appendix~\ref{appendix:CII_comparison}, we use a simple ionisation calculation to confirm that our model is consistent with the \ion{C}{2} emission detected by \citet{Donaldson_etal_2013}.

\begin{figure}
\plotone{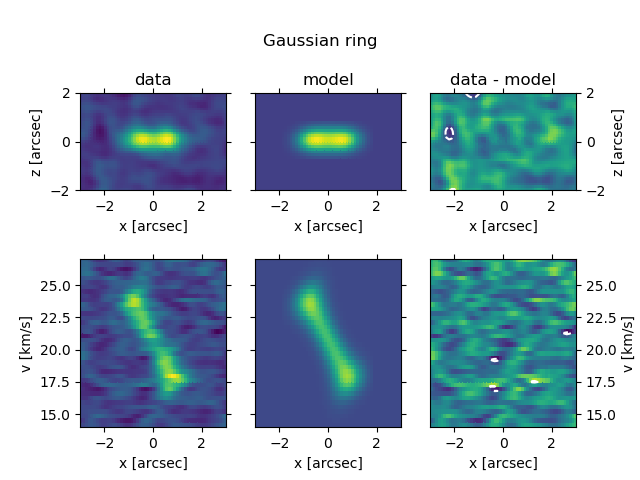}
\caption{The Gaussian ring model with the highest posterior probability. The top row shows the moment 0 and the bottom row the pv diagram. The columns show the data, the model and the residual (data minus model, with contours in steps of 3$\sigma$) respectively. 
The parameter values for this model are as follows: $i=77.4$\arcdeg, $r_0=108$\,au, $\sigma_r=36$\,au, $T_{100}=76$\,K, $M_\mathrm{C^0}=3.4\times10^{-3}$\,M$_\oplus$, $M_*=1.6$\,M$_\odot$, $v_*=20.7$\,km\,s$^{-1}$, $\Delta x=-0.06$\arcsec, $\Delta y=0.04$\arcsec.\label{fig:highest_prop_Gaussian_ring}}
\end{figure}

\begin{table*}
\centering
\caption{Fitted parameters for the Gaussian ring model (section \ref{sec:CI_modeling}) and the double power law model (Appendix~\ref{appendix:power_law_models}). The parameters $\beta_\mathrm{in}$ and $\beta_\mathrm{out}$ are restricted to the double power law model.}\label{tab:fit_parameters}
\begin{tabular}{cccclc}
\hline
\hline
Parameter & \multicolumn{2}{c}{value\tablenotemark{a}} & unit & prior\tablenotemark{b} & range\tablenotemark{c}\\
&Gaussian & DPL\tablenotemark{d}& & & \\
\hline
$i$ & $77.9^{+1.7}_{-1.5}$& $78.0^{+1.8}_{-1.5}$ & deg & loc & [65,90] \\
$r_0$ & $109^{+3}_{-4}$ & $129^{+8}_{-11}$ & au & loc & [30,200] \\
$\sigma_r$ & $35.4^{+3.8}_{-3.5}$ & - & au & sca & [0,100] \\
$\beta_\mathrm{in}$ & - & $1.5^{+1.1}_{-0.6}$  & - & loc & [-10,10] \\
$\beta_\mathrm{out}$ & - & $-7.4^{+1.5}_{-1.6}$  & - & loc & [-10,10] \\
$T_{100}$ & $117^{+98}_{-60}$ & $125^{+101}_{-64}$ & K & sca & [10,300] \\
$M_\mathrm{C^0}$ & $3.5^{+0.2}_{-0.2}$ & $3.6^{+0.2}_{-0.2}$ & 10$^{-3}$\,M$_\oplus$ & sca & [0,1000] \\
$M_*$ & $1.59^{+0.05}_{-0.05}$ & $1.59^{+0.05}_{-0.05}$ & M$_\odot$ & sca & [1,3] \\
$v_*$ & $20.67^{+0.04}_{-0.04}$ & $20.67^{+0.04}_{-0.04}$ & km\,s$^{-1}$ & loc & [18,22] \\
$\Delta x$ & $-0.06^{+0.01}_{-0.01}$ & $-0.06^{+0.01}_{-0.01}$ & arcsec & loc & [-0.75,0.75] \\
$\Delta z$ & $0.04^{+0.01}_{-0.01}$ & $0.04^{+0.01}_{-0.01}$ & arcsec & loc & [-0.75,0.75] \\
\hline
\end{tabular}
\tablenotetext{a}{50th percentile with error bars corresponding to the 16th and 84th percentile of the posterior distribution.}
\tablenotetext{b}{Location invariant (loc) or scale invariant (sca) prior.}
\tablenotetext{c}{Parameter range explored by the MCMC fit.}
\tablenotetext{d}{Double power law}
\end{table*}

\section{Model for gas production and destruction} \label{sec:gas_production_model}

With the observations of CO \citep{MacGregor_etal_2018,Moor_etal_2019} and carbon (this work) at hand, we can now attempt to model the history of gas production. In the simplest story, two parameters describe this history: the event time, when gas production began, and the CO gas production rate. The value for the event time could be compared with the system age to infer something about the origin of the debris disk, and the CO production rate might be compared with the dust production rate to constrain the composition of the planetesimals. Naively, one expects the CO production rate to be a fraction of the dust mass loss rate \citep[5.2\,M$_\oplus$\,Myr$^{-1}$,][]{Moor_etal_2019}, reflecting the fractional CO content of the gas-producing bodies \citep[typically assumed to be $\sim$10\%, as suggested by outgassing solar system comets,][]{Huebner_2002}. Our detailed modelling below suggests that a significantly higher CO production rate is required to explain the observed C and CO masses.

By using an ATLAS9 model\footnote{The young A-star $\beta$~Pic is observed to emit above a standard model atmosphere at wavelengths $\sim$900--1100\,\r{A} relevant for CO photodissociation \citep{Deleuil_etal_2001,Bouret_etal_2002,Roberge_etal_2006}. Could the same be true for HD~32297? Unfortunately, to the best of our knowledge, no data in this wavelength region are available for HD\,32297. However, we inspected HST/STIS spectra at longer wavelengths (1150--1700\,\r{A}) and did not find evidence for emission above a standard atmospheric model.} tailored to HD~32297, we find that the CO photodissociation and C ionisation rates are dominated by the ISRF beyond $\sim$12\,au from the star. This allows us to ignore stellar radiation.

To put the following model into context, we first make a simple order-of-magnitude estimate for the event time. We assume that CO molecules are photo-dissociated in the two surface layers of the disk by the ISRF without abatement, down to a column density of $N_\mathrm{CO}=10^{15}$\,cm$^{-2}$ \citep[behind such a column density, the dissociation rate is reduced by a factor 0.4 due to self-shielding,][]{Visser_etal_2009}. This implies a total CO destruction rate of $D_{\rm CO}=2\pi r_0\Delta r(2N_\mathrm{CO})m_\mathrm{CO}/(240\mathrm{yr}) \sim 0.8\,M_\oplus/$Myr (where $m_\mathrm{CO}$ is the mass of a CO molecule, $r_0 = 110$\,au and $\Delta r = 80$\,au are from the the Gaussian ring model, (table~\ref{tab:fit_parameters}) and the CO lifetime is 240\,yr, i.e.\ twice the unshielded lifetime since half of the ISRF is blocked by the disk for molecules on the surface). To accumulate the observed C$^0$ mass of $M_{\mathrm{C}^0} \approx 3.5\times 10^{-3}$\,M$_\oplus$ (table \ref{tab:fit_parameters}) will require a timescale of 
\begin{equation}
    t_0 \approx \frac{M_{\mathrm{C}^0}}{(12/28)D_{\rm CO}} \sim 10^4\,{\rm yr}.
\end{equation}
Compared to the system lifetime, this is a surprisingly short timescale. As our model below shows, additional shielding of CO by carbon is unable to explain the short duration of the production time we derive here. Instead, the removal of carbon atomic gas by dust grains might alleviate the discrepancy between a longer production time and the low observed carbon mass.

\subsection{Time Evolution Model}\label{sec:time_evolution_model}
Here, we present a simple model to capture a few main processes that affect the time evolution of CO and carbon. This includes UV shielding of CO by carbon, CO self-shielding and carbon removal (e.g.\ via re-capture by dust grains). The shielding aspect of our model is similar in spirit to that in \citet{Kral_etal_2019,Moor_etal_2019,Marino_etal_2020}, but differs (substantially) in the treatment of shielding. 

We consider only one radial zone of well-mixed gases and discard the effect of gas removal by viscous spreading (as is justified here, see section \ref{sec:other_C_removal_processes}).
The CO mass evolves as
\begin{equation}\label{eq:CO_mass_evolution}
    \frac{dM_{\rm CO}}{dt} = P_\mathrm{CO} - D_\mathrm{CO},
\end{equation}
where $P_\mathrm{CO}$ is the CO production rate by the debris ring and $D_\mathrm{CO}$ the CO photo-destruction rate. CO is produced in a collisional cascade among the debris\footnote{We neglect CO production by the gas phase reaction C+O$\rightarrow$CO+$\gamma$. The rate coefficient is $k=2.1\times10^{-19}$\,cm$^3$\,s$^{-1}$ \citep[at $T\leq300$\,K,][]{Glover_etal_2010}. For typical densities, this results in negligible CO production.}. Intuitively, we might expect some time dependence of $P_\mathrm{CO}$ such as a gradual decrease due to the collisional evolution of the disk \citep{Marino_etal_2020}. However, for simplicity, we assume a constant CO production rate. The carbon mass evolves with time as
\begin{equation}\label{eq:C_mass_evolution}
    \frac{dM_{\rm C}}{dt} = \frac{12}{28} D_{\rm CO} - \frac{M_\mathrm{C}}{\tau_\mathrm{C}}
\end{equation}
The second term on the right-hand-side models a carbon sink (e.g.\ capture onto dust grains). To convert $M_\mathrm{C}$ into masses of neutral and ionised carbon, a constant ionisation fraction of 0.2 is assumed (appendix \ref{appendix:CII_comparison}).

In contrast to \citet{Kral_etal_2019}, we obtain the photo-dissociation rate ($D_\mathrm{CO}$) using photon counting. The probability that a given ISRF photon either ionises C$^0$ or dissociates CO is given by
\begin{equation}\label{eq:ISRF_photon_interaction_probability}
f_1=1-\exp(-\tau_\mathrm{C^0}-\tau_\mathrm{CO})=1-\exp(-\tau_\mathrm{tot})
\end{equation}
where the optical depth $\tau_x(\lambda) = N_x \sigma_x(\lambda)$ with $N_x$ the vertical column density and $\sigma_x$ the wavelength-dependent cross section. Under our assumption of perfect mixing, the fraction of these photons dissociating CO is
\begin{equation}
f_2 = \frac{n_\mathrm{CO}\sigma_\mathrm{CO}}{n_\mathrm{C^0}\sigma_\mathrm{C^0}+n_\mathrm{CO}\sigma_\mathrm{CO}}=\frac{\tau_\mathrm{CO}}{\tau_\mathrm{tot}}
\end{equation}
where $n$ denotes number density. Let $\phi_\nu$ be the number of ISRF photons hitting the disk per unit time, per unit frequency, per unit area \citep[taken from][]{Draine_1978}. The photo-destruction mass rate for CO is 
\begin{equation}\label{eq:CO_destruction_rate}
    D_\mathrm{CO} = m_{\rm CO} \int\phi_\nu A f_1 f_2 \, d\nu
\end{equation}
where $A=2\pi r_0\Delta r$ is the surface of the disk. An alternative derivation of this equation is given in appendix \ref{appendix:alternative_CO_destruction_rate_derivation}.

Equation \ref{eq:CO_destruction_rate} can be used to derive an upper limit on the CO destruction rate. We assume $f_1=f_2=1$ for any frequency $\nu$ where $\sigma_\mathrm{CO}(\nu)>0$, i.e., no photon that could potentially destroy a CO molecule is lost. This leads to $D_\mathrm{CO}^\mathrm{max}\sim50$\,M$_\oplus$\,Myr$^{-1}$ (with $r_0=110$\,au and $\Delta r$=80\,au). In the case that  $P_\mathrm{CO}>D_\mathrm{CO}^\mathrm{max}$, no steady state is possible.

We extend our model to also include the evolution of $^{13}$CO and C$^{18}$O. This is easily achieved by replacing $\tau_\mathrm{CO}$ in equation \ref{eq:ISRF_photon_interaction_probability} by $\tau_\mathrm{^{12}CO}+\tau_\mathrm{^{13}CO}+\tau_\mathrm{C^{18}O}$. Then, the masses of the isotopologues are evolved the same way as $^{12}$CO. The wavelength-dependent cross sections of the different isotopologues only overlap partially. Therefore, the lower abundance isotopologues are more readily photodissociated since their self-shielding is weaker. This effect is called isotopologue selective photodissociation. It increases the ratios of $^{12}$CO/$^{13}$CO and $^{12}$CO/C$^{18}$O compared to the ISM ratios, and is naturally taken into account by our model.

The cross-sections for CO photo-dissociation were provided by A.\ Heays (private communication) and are based on \citet{Visser_etal_2009}, while the ionization cross section $\sigma_\mathrm{C}$ is from the database of 'Photodissociation and photoionization of astrophysically relevant molecules'\footnote{\url{https://home.strw.leidenuniv.nl/~ewine/photo/}} \citep[hereafter PPARM,][]{Heays_etal_2017}; $\sigma_\mathrm{C}$ is essentially constant ($\sim$1.6$\times10^{-17}$\,cm$^2$) over the wavelength range where CO photodissocation happens \citep{Rollins_Rawlings_2012}.

We note that the \citet{Kral_etal_2019} model has a different approach to the calculation of the CO destruction rate and we argue that they under-estimate it in the shielded regime. In their model,
\begin{equation}\label{eq:Kral_CO_destruction_rate}
D_\mathrm{CO}=M_{\rm CO}\times \chi_0\times S(N_\mathrm{CO}) \times \exp(-N_\mathrm{C^0}\sigma_\mathrm{C^0})
\end{equation}
where $\chi_0$ is the unshielded ISRF photodissociation rate and $S(N_{\rm CO})$ is the reduction on $\chi_0$ due to CO self-shielding under a column density $N_{\rm CO}$ \citep{Visser_etal_2009}. However, equation \ref{eq:Kral_CO_destruction_rate} is the dissociation rate for a CO molecule that sits behind the full column densities of C and CO. First, such a model ignores the CO dissociation in the shielding layer and therefore over-estimates the effect of shielding. In contrast, our model predicts higher CO destruction rates because even if the mid-plane is strongly self-shielded, ISRF photons are still used to dissociate CO in the higher layers of the disk. Second, the \citet{Kral_etal_2019} model also implicitly assumes that C forms a separate layer at the surface of the disk, as expressed by the exponential term in equation \ref{eq:Kral_CO_destruction_rate}. This maximises the effect of C shielding. In contrast, in our model, C and CO are well-mixed. Observations of \ion{C}{1} that resolve the vertical structure are needed to decide which prescription is a better representation of a real disk.

Next, we consider carbon condensation onto dust grains. We estimate the total dust cross section by adopting a dust mass of 0.57\,M$_\oplus$ \citep{MacGregor_etal_2018}, a minimum grain diameter equal to the blowout limit ($\sim$10\,$\mu$m), a maximum grain diameter of 1\,mm and a power-law size distribution with index -3.5. Adopting $T=100$\,K and $n_\mathrm{C}=370$\,cm$^{-3}$ (from our Gaussian ring model fit), we find that the mean free time of a C atom before colliding with a dust grain is $\sim$5000\,yr. \citet{Grigorieva_etal_2007} also estimated the recondensation of water onto dust grains. By adapting their equation 17\footnote{We note an error in that equation. The numerical pre-factor should be $1.8\times10^{-20}$ rather than $1.8\times10^{-26}$.} to C, we find that all C is recondensed on a timescale of $\sim$8000\,yr. So this process might be important. Although collisions between C atoms and the grains have nearly unity sticking coefficient \citep[e.g.][]{Hollenbach_Salpeter_1970,Leitch-Devlin_Williams_1985}, thermal desorption might remove them again quickly. In the later discussion, we consider this more in detail and also discuss whether the accreted atoms can form fresh CO (section \ref{sec:reformation}).

\begin{figure}
\plotone{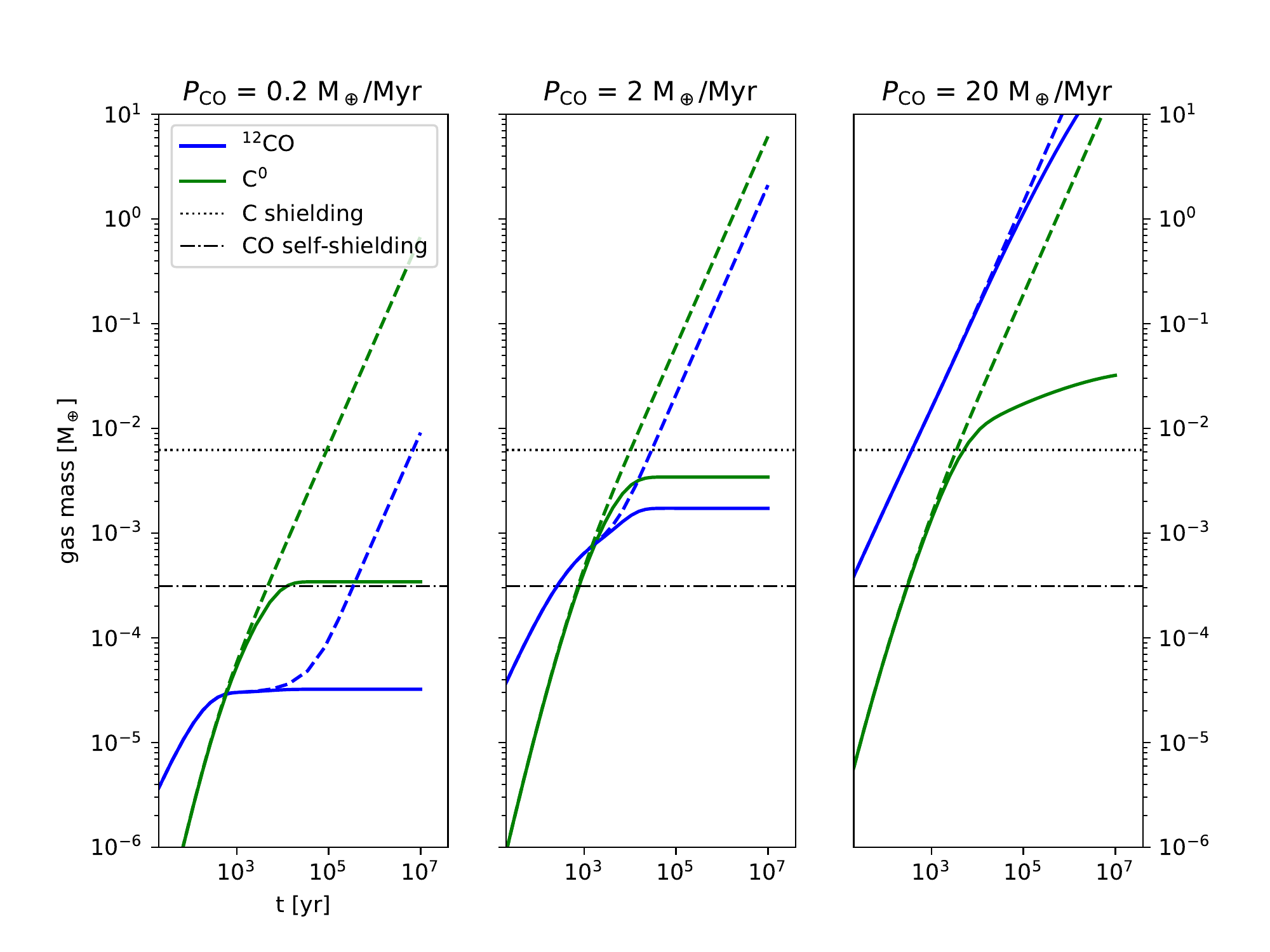}
\caption{Evolution of the CO and C$^0$ mass for different CO production rates (rising from the left to the right). The solid lines are for models with $\tau_\mathrm{C}=5000$\,yr and the dashed lines for models without C removal ($\tau_\mathrm{C}=\infty$). The horizontal lines indicate the masses at which C$^0$ shielding (dotted) or CO self-shielding (dash-dotted) become significant, i.e.\ the dissociation rate in the midplane is reduced by a factor 0.3. For models with $\tau_\mathrm{C}=\infty$, shielding allows both masses to rise linearly with time, and the mass ratio of C$^0$/CO remains constant. In contrast, models with C removal settle to a steady state within the simulated time span, except for the highest CO production rate considered here (right panel).
\label{fig:typical_gas_evolution}}
\end{figure}

In the following, we leave the removal timescale $\tau_\mathrm{C}$ as a free parameter in our model. Figure~\ref{fig:typical_gas_evolution} illustrates three examples of our model for different rates of CO production without and with removal of carbon (taking $\tau_\mathrm{C} = \infty$ and 5000\,yr, respectively). For these models, we fixed $r_0=110$\,au and $\Delta r = 80$\,au. We find that without C removal, once either C$^0$ shielding or CO self-shielding is significant, both C and CO rise linearly and the C$^0$/CO ratio remains constant. When the CO production rate is low, C shielding is necessary before CO can survive rapid destruction. This explains the initial delay in its mass rise (left panel). Later on, both masses grow linearly, with a fixed fraction of the freshly made CO being turned into C. The ratio of C$^0$/CO is above unity. In comparison, in the models with a high CO production rate (middle and right panels), even from the very beginning, CO self-shielding is already significant. We still observe a linear growth in both masses, but the C$^0$/CO mass ratio is consistently below unity. On the other hand, removal of C gas allows the system to settle to a steady state in which the removal of C is balanced by its production from CO dissociation. We note that the C$^0$/CO ratio decreases with increasing CO production rate, regardless of whether C is removed or not. For HD~32297, the observed C$^0$/CO$\ll1$ suggests that a large CO production rate will be required to explain the observations.
 
\subsection{Applying to HD\,32297}
Here, we apply our above model to the specific case of HD\,32297. The planetesimal belt is observed to have a width of $\sim$40\,au if we ignore the 'halo component' \citep{MacGregor_etal_2018}. Thus, the C gas with its FWHM of $\sim$80\,au is observed to be mildly more spread out. For simplicity, we ignore this difference and fix the width of the gas disk to 80\,au.

Our model contains two unknowns: the event time ($t_0$) at which CO (and presumably dust) production commenced, and the CO production rate ($P_{\rm CO}$). We aim to reproduce two observables: the C$^0$ mass ($3.5\times 10^{-3}$\,M$_\oplus$) and the mass of C$^{18}$O \citep[$1.4\times10^{-4}$\,M$_\oplus$,][]{Moor_etal_2019}. We use the C$^{18}$O mass because the main component of the CO gas, $^{12}$CO, is optically thick and its mass cannot be determined accurately. We also ignore $^{13}$CO as it may be optically thick as well.

\subsubsection{Steady-state solution}\label{sec:steady_state_solution}

We first consider the case where the system described by equations \ref{eq:CO_mass_evolution} and \ref{eq:C_mass_evolution} is in a steady state. A steady state solution only exists if C is removed, i.e.\ $\tau_\mathrm{C}$ is finite. 

In steady state, destruction rates equal production rates: $P_\mathrm{^yC^zO}=D_\mathrm{^yC^zO}$ with $^y$C$^z$O a specific CO isotopologue. The destruction rate $D_\mathrm{^yC^zO}$ can be calculated if masses of all CO isotopologues and of C$^0$ are known (equation \ref{eq:CO_destruction_rate}). However, as mentioned above, only the C$^{18}$O mass is well measured \citep{Moor_etal_2019}. To proceed, we first compute the $^{12}$CO and $^{13}$CO masses using the measured C$^{18}$O mass, assuming ISM ratios: $\mathrm{^{13}CO}/\mathrm{^{12}CO}=1.4\times10^{-2}$ and $\mathrm{C^{18}O}/\mathrm{^{12}CO}=1.8\times10^{-3}$ \citep{Wilson_1999}. We then calculate the destruction rate of each isotopologue using equation \ref{eq:CO_destruction_rate}. By the steady state assumption, these destruction rates are equal to the production rates. We find that the production rates of $^{13}$CO and C$^{18}$O (relative to $^{12}$CO) are enhanced by a factor $\sim$6 relative to the ISM ratios. This would imply that the gas-producing solids are enhanced in $^{13}$CO and C$^{18}$O. However, isotopologue ratios in solar system comets show $^{12}$C/$^{13}$C and $^{16}$O/$^{18}$O ratios broadly similar to the ISM ratios \citep[e.g.][and references therein]{Hassig_etal_2017,Altwegg_etal_2019}. This motivates us to search for values of the $^{12}$CO and $^{13}$CO masses that imply CO production rate ratios consistent with ISM ratios. We find $M_{^{12}\mathrm{CO}}=1.6$\,M$_\oplus$ and $M_{^{13}\mathrm{CO}}=1.5\times10^{-3}$\,M$_\oplus$, i.e., we require a $^{12}$CO/C$^{18}$O mass ratio enhancement by a factor 20 relative to the ISM value. This is not unexpected, because isotopologue selective photodissociation can preferentially destroy the rarer isotopologues. The CO production rate inferred from these masses is very high, $16$\,M$_\oplus$\,Myr$^{-1}$. The corresponding steady state C removal timescale (equation \ref{eq:C_mass_evolution}) is short: $\tau_\mathrm{C}\approx 700$\,yr. This is a consequence of the small C$^0$ mass we measure.

We note that with a CO mass as high as $1.6$\,M$_\oplus$, carbon shielding plays a relatively minor role---when neglecting the shielding contribution of neutral carbon, the CO destruction rate is increased by only $\sim$20\%.

\subsubsection{Full calculation}

We now do not assume steady state and instead evolve a grid of models with different event times ($t_0$) and CO production rates ($P_{\rm CO}$) with the goal of reproducing the observed masses of C$^0$ and C$^{18}$O. We fix the isotopologue ratios in the gas-producing solids (i.e.\ the CO production rate ratios) to ISM ratios.

\begin{figure}
\plotone{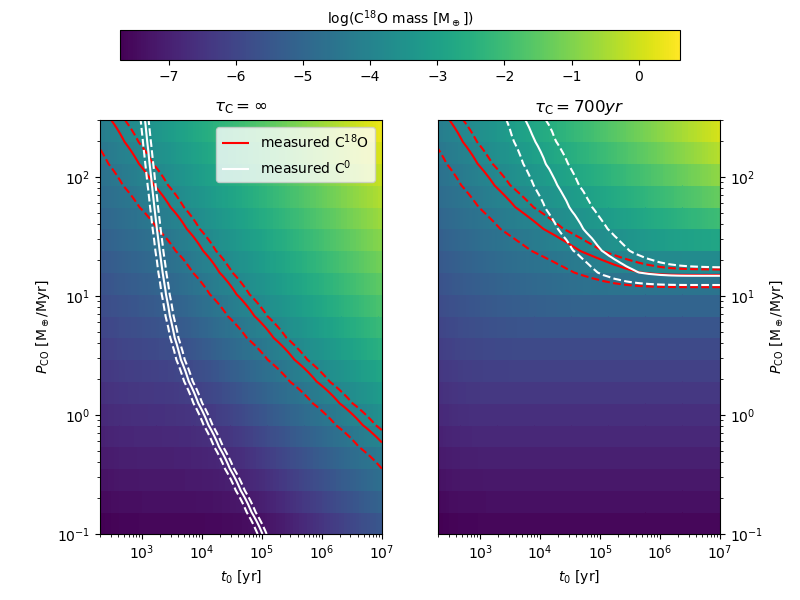}
\caption{C$^{18}$O mass (with color bar) calculated from our model as a function of CO production rate (vertical axis) and event time (horizontal axis) for a scenario without removal of C (left) and with $\tau_\mathrm{C}=700$\,yr (right). The solid red and white lines show the parameter combinations that give rise to the observed C$^{18}$O \citep[$1.4\times10^{-4}$\,M$_\oplus$,][]{Moor_etal_2019} and C$^0$ ($3.5\times10^{-3}$\,M$_\oplus$, this work) masses respectively, with the dashed lines corresponding to the $\pm3\sigma$ uncertainty ranges. The correct parameters lie where these two sets of lines intersect. In the right panel, evolution settles down to a steady state, so we lose sensitivity to the event time after $\sim 10^5$\,yr.
\label{fig:PCO_vs_tend}}
\end{figure}

Our model results are presented in Figure~\ref{fig:PCO_vs_tend}. In the case of no C removal ($\tau_\mathrm{C}=\infty$, left panel), we find that only the combination of an extremely high CO production rate and a recent event can explain the data: $P_\mathrm{CO}\approx100$\,M$_\oplus$\,Myr$^{-1}$ and $t_0\approx 1000$\,yr. As can be seen from the figure, for lower CO production rates, that model exceeds the observed C mass long before it has built up the observed C$^{18}$O mass.

Figure \ref{fig:PCO_vs_tend} (right) shows a scenario were $\tau_\mathrm{C}$ is adjusted such that the observed gas masses can be explained with a steady state solution. We find $\tau_\mathrm{C}\sim700$\,yr is necessary, i.e.\ C gas needs to be removed efficiently in order to reproduce the small C mass we observe: for larger values of $\tau_\mathrm{C}$, the results are qualitatively similar to the $\tau_\mathrm{C}=\infty$ case, while for smaller values, no solution is found. Even in this scenario, the required CO production rate remains high at $\sim$15\,M$_\oplus$\,Myr$^{-1}$ (as mentioned in section \ref{sec:gas_production_model}, based on the dust mass loss rate, we would expect a CO production rate of $\sim$0.5\,M$_\oplus$\,Myr$^{-1}$). Steady state is reached after a few times $10^5$ years. The steady state CO mass is 1.3\,M$_\oplus$. These results are consistent with the steady state solution described in section \ref{sec:steady_state_solution}.

\begin{figure}
\plotone{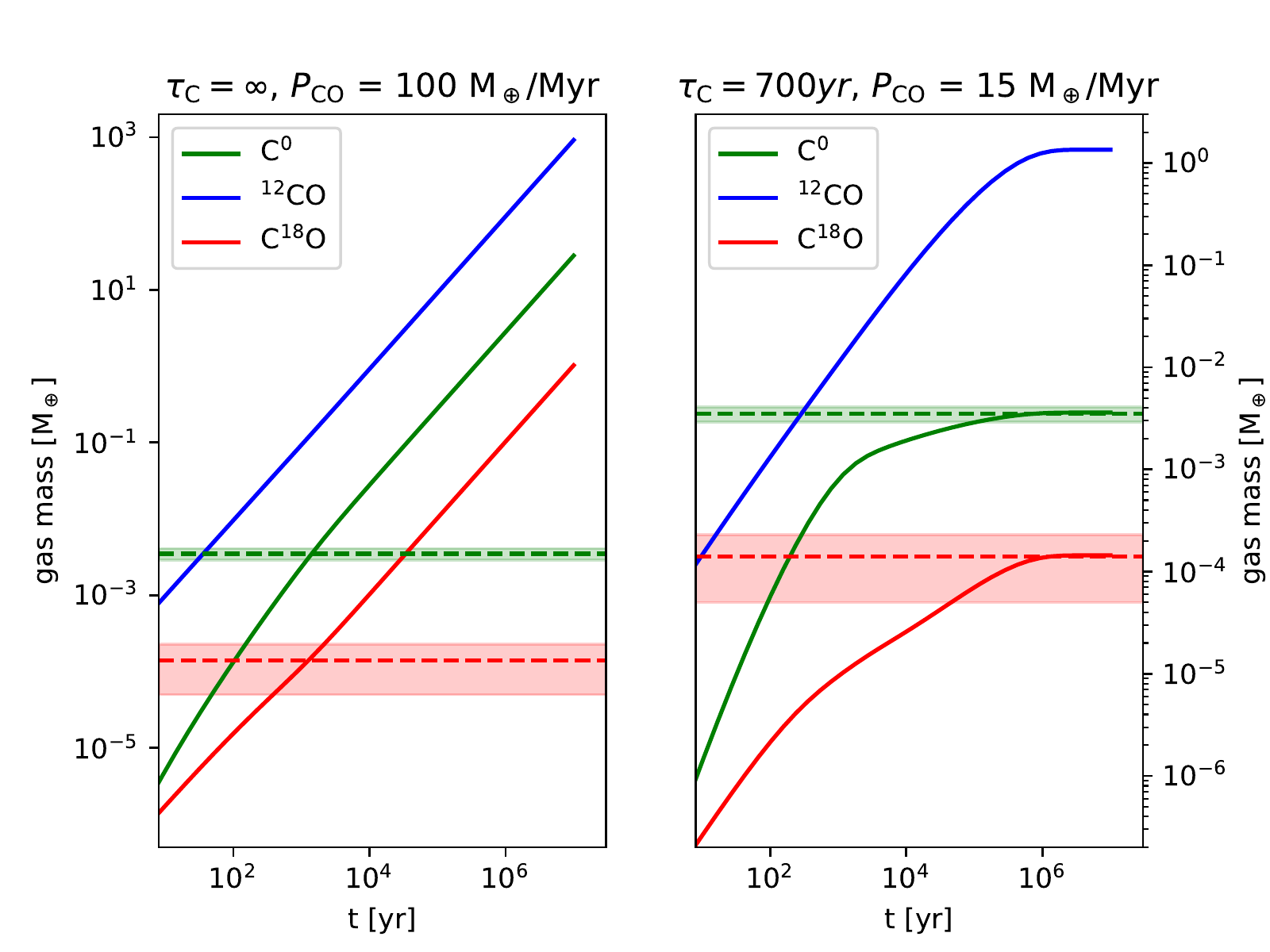}
\caption{Evolution of models without (left) and with (right) removal of C gas that can reproduce the observed C$^0$ and C$^{18}$O masses (model parameters as inferred from figure \ref{fig:PCO_vs_tend}). The horizontal dashed lines indicate the observed masses, with the shaded regions corresponding to $\pm3\sigma$ ranges.}
\label{fig:gas_evolution_fitting_models} 
\end{figure}

In Figure \ref{fig:gas_evolution_fitting_models}, we show the two models that can formally explain the observed gas masses, with parameters as identified from Figure \ref{fig:PCO_vs_tend}. As mentioned above, the first model (no removal of carbon) requires a very recent event and an extremely large CO production rate. By the time the model fits the observed C$^0$ and C$^{18}$O masses, the total CO mass released is $\sim$0.1\,M$_\oplus$. This model can thus be interpreted as the recent destruction of a $\sim$1\,M$_\oplus$ body with a CO fraction of 10\% releasing all of its CO within 1000\,yr. In the following, we focus on discussing the second model. In this latter case, the model does not give us a unique answer regarding the event time, other than that it must be greater than $\sim 10^5$\,yr. For comparison, the system lifetime is larger than 15\,Myr. The CO production rate of 15\,M$_\oplus$\,Myr$^{-1}$ is in contrast with the result from \citet{Moor_etal_2019}, where they found that $P_\mathrm{CO}\approx3.6\times10^{-2}$\,M$_\oplus$\,Myr$^{-1}$. However, the \citet{Moor_etal_2019} model over-predicts the C$^0$ mass by a factor 10 compared to our data, despite underestimating the CO break-down rate due to the different treatment of shielding (section \ref{sec:time_evolution_model}).
 
\section{Discussion}\label{sec:discussion}

The results of our steady state model have two obvious contradictions: the inferred CO production rate (15\,M$_\oplus$\,Myr$^{-1}$) exceeds the total dust mass loss rate  \citep[5.2\,M$_\oplus$\,Myr$^{-1}$;][]{Moor_etal_2019}; and the C removal timescale ($\tau_C = 700$ yr) is shorter than what we naively estimated based on dust cross-section ($\sim$5000\,yr, see section \ref{sec:time_evolution_model}).

For a CO (and CO$_2$) mass fraction of $10\%$, we expect a CO production rate $\sim$30 times below what we obtain above. A number of factors may bring some relief: uncertainties on the value of dust grind-down rate can be up to a factor of 10 \citep{Kral_etal_2019}; the CO production rate may be time-dependent, as opposed to being constant as assumed here \citep{Marino_etal_2020}. But a tension remains in that our required CO production rate is higher than the total dust rate.

In the following, we discuss a new process: CO re-formation. We describe how this process could allow the CO production rate to be (much) larger than the dust grind-down rate. It could also  explain the belt-like appearance of the gas disk. We then go on to discuss the inferred carbon removal time and a number of alternative removal processes. We end by discussing the implications of our results for the origin of debris disks.

\subsection{Carbon removal by dust grains and CO re-formation}\label{sec:reformation}
Here, we consider the fate of carbon atoms that collide with dust grains. We also discuss the possibility that accreted C atoms will react with O to reform CO. Although \ion{O}{1} 63\,$\mu$m emission was not detected by \textit{Herschel} \citep{Donaldson_etal_2013}, we expect O to be at least as abundant as C, thanks to photodissociation of CO, CO$_2$ and H$_2$O, all of which should be present in the cometary parent bodies.

We first consider if atoms can remain on the grain or are quickly re-relased to the gas phase by thermal desorption. The rate of thermal desorption $R_\mathrm{td}$ depends on the binding energy $E_i$ \citep{Hollenbach_Salpeter_1970,Tielens_Hagen_1982}:
\begin{equation}\label{eq:thermaldesorp}
R_{\rm td} \approx \nu_i e^{- E_d/k T_{\rm gr}}
\end{equation}
where $\nu_i\sim 10^{13}$\,s$^{-1}$ \citep{Tielens_Hagen_1982} is the characteristic oscillation frequency for the atom in the lattice potential. The value of the binding energy depends on the type of interaction between the adsorbate and the surface \citep[e.g.][]{Tielens_2005}. Binding by van der Waals forces (physiosorption) is relatively weak \citep[0.01--0.2\,eV,][]{Tielens_2005}, while valence bonds (chemisorption) are stronger ($\sim$1\,eV). Whether adsorbed atoms are physiosorbed or chemisorbed depends on the grain properties (e.g.\ chemical composition, temperature; Y.\ Aikawa, private communication). For instance, let us assume a grain temperature of 50\,K. If C is physiosorbed with a binding energy of 0.1\,eV, it would be thermally desorbed in a fraction of a second. In such a case, dust grains cannot act as C sink. On the other hand, chemisorbed C can be accumulated on the grains. More studies \citep[e.g.\ quantum chemical computations such as in][]{Shimonishi_etal_2018} are needed to determine the binding energy of C (and O) on dust grains.

Next, we consider the possibility that accreted C and O atoms reform CO on the dust grains. Detailed modelling to determine if and under which conditions C+O$\rightarrow$CO can proceed on the dust grains\footnote{Although the formation of CO is energetically favorable (the CO binding energy is $\sim 10$\,eV), the reaction might have a barrier if C and/or O are strongly bound to the dust grain.} is beyond the scope of this paper. However, if the reaction proceeds, it has interesting consequences. CO is then continuously recycled: it is released from grains, photo-dissociated by the ISRF, C and O re-accreted onto grains, CO re-formed rapidly, and re-released into space. In this picture, as long a certain amount of CO is injected as gas at some point, there is no need for fresh CO production. The CO production rate that we infer would simply be a result of CO reformation. Consider a C mass of $3\times 10^{-3}$\,M$_\oplus$ and a removal time of $\tau_\mathrm{C} = 700$\,yr. This yields the rate of CO reformation, $\sim 10$\,M$_\oplus$\,Myr$^{-1}$, as we obtain above. It is not necessary for the CO production rate to be limited by the dust grind-down rate. CO re-formation would also explain why the carbon gas is concentrated into a narrow radial belt---all gas components have been released so recently that they did not have time to spread viscously.

We experimented with a modified version of our model with an additional CO source term equal to $28/12M_\mathrm{C}/\tau_\mathrm{C}$, i.e.\ each removed C atom is immediately converted to a CO molecule. Then, the observed C and CO masses can indeed be reproduced with much smaller CO production rates ($P_\mathrm{CO}\lesssim1$\,M$_\oplus$\,Myr$^{-1}$) compared to our standard model.

\subsection{Other C removal processes}\label{sec:other_C_removal_processes}

While consideration of dust capture yields a carbon removal time of $\tau_C \sim 5000$\,yr (section \ref{sec:time_evolution_model}), our model requires a much shorter $\tau_c \sim 700$\,yr, in order to explain the low C$^0$ mass we infer from the ALMA observations. We consider a number of alternative carbon removal mechanisms, but do not find a satisfactory solution to this problem at the moment.
\begin{enumerate}
\item Gas accretion into the star appears ineffective in HD~32297, as the gas ring appears narrow, not radially spread into an accretion disk. Indeed, our modeling of the \ion{C}{1} emission suggests that the gas is distributed in a ring that is only mildly broader than the dust ring.

\item Can one or several planets inward of the dust belt prevent the formation of an accretion disk by accreting gas themselves? So far, no giant planets have been found in direct imaging searches around HD\,32297 \citep{Boccaletti_etal_2012,Rodigas_etal_2014,Bhowmik_etal_2019}. In the most recent study, \citet{Bhowmik_etal_2019} exclude the presence of planets with 2.5--6 Jupiter masses at projected distances beyond $\sim$80\,au. However, because of the edge-on geometry of the disk, planets might be hiding close to the line of sight towards the star. Furthermore, \citet{Marino_etal_2020} found that smaller, currently undetectable Neptune-class planets can be sufficient to accrete gas.

\item \citet{Marino_etal_2020} recently suggested that C might become undetectable as it viscously spreads outwards to regions at larger radii with lower surface density. More detailed modeling is needed to test this proposition. Naively, one might expect this to be less of a problem for an edge-on disk like HD~32297 where the column density along the line of sight is high compared to a face-on disk. This scenario would also imply efficient spreading and thus would again require a mechanism to prevent the formation of an inward accretion disk (see the previous point).

\item What about carbon removal by gas phase chemistry? Consulting the chemical network and reaction rates relevant for proto-planetary disks, as compiled by \citet{Glover_etal_2010}, we find that the most probable pathway for carbon removal is the following reaction\footnote{Another reaction, $\mathrm{C}+\mathrm{O}_2 \rightarrow \mathrm{CO}^+ + \mathrm{O}$, is likely less important on account of the lower density of the fragile O$_2$ molecules.}
\begin{equation}\label{eq:OH}
    \mathrm{OH} + \mathrm{C} \rightarrow \mathrm{CO} + \mathrm{H}, 
\end{equation}
with a reaction rate of $k = 10^{-10}$\,cm$^3$\,s$^{-1}$ (similarly, OH + C$^+$ $\rightarrow$ CO$^+$ + H has $k=7.7\times10^{-10}$\,cm$^3$\,s$^{-1}$). Thus, to match our required C removal timscale of $\tau_\mathrm{C}=700$\,yr, we would require $n_\mathrm{OH}=(\tau_\mathrm{C}k)^{-1}=0.5$\,cm$^{-3}$. The OH molecules are likely the results of water photo-dissociation, and they are highly vulnerable to UV photo-dissociation. So for this channel to be effective in removing carbon gas, one requires a copious amount of water production. We encourage more detailed modeling to assess how chemistry influences the evolution of debris disk gas.

\item Can carbon gas spontaneously condense into solid form, without the assistance of dust grains? The low-temperature phase diagram of carbon \citep[see e.g.][]{Jaworski_etal_2016} shows that the vapour pressure for condensation of gaseous atomic C into solid state at the relevant temperature range is orders of magnitude below the gas pressure in the disk. However, typical condensation requires the presence of seed nuclei. These nuclei ought to be scarce as they are quickly removed by the radiation pressure from the A-star \citep{Weingartner_Draine_2001}, although sufficient amounts of gas could lengthen their residence time.
\end{enumerate}

Another possible solution might be to assume a CO production that is enhanced in isotopologues. For instance, we experimented with a model where the production rate of $^{13}$CO and C$^{18}$O, relative to $^{12}$CO, is enhanced by a factor of 10 from the ISM ratio. At steady state, we require a CO production rate of 5\,M$_\oplus$\,Myr$^{-1}$, and a carbon removal time of $\tau_c \sim 2000$\,yr, more similar to the 5000\,yr inferred in section \ref{sec:time_evolution_model}.

\subsection{Other Uncertainties}\label{sec:other_model_uncertainties}

The conclusions we draw from our modelling rely heavily on our main observational result: the low mass of neutral carbon. How robust is this mass measurement? Could it be that the small error bars (see Table~\ref{tab:fit_parameters}) derived from the MCMC modelling are underestimating the real uncertainty? One scenario that could increase the C$^0$ mass estimate is a low kinetic temperature. This could lead to a high optical depth, allowing for an arbitrarily high C$^0$ mass, in principle. We discuss this possibility in appendix \ref{appendix:mass_uncertainty_low_Tkin}. Overall, we believe that our mass error-bars are reasonable.

Regarding UV shielding, we can exclude the possibility of another shielding agent. Other than carbon and CO, there are no promising molecules or atoms in the PPARM database that have the required high cross-section at the relevant wavelengths to provide shielding.

Next, the strength of the ISRF may be somewhat uncertain. Very few direct observations exist \citep[e.g.][]{Henry_2002}, and there may also be some spatial variations caused by proximity to massive stars \citep[e.g.][]{vanDishoeck_1994}. A weaker ISRF than the canonical value assumed here would reduce the CO destruction rate, and correspondingly, the inferred CO production rate.

Finally, our model assumes that CO and C are well-mixed. When instead assuming that all C is concentrated at the surface of the disk (this configuration maximises C shielding), the derived CO production rates and C removal timescales only change by a factor of $\sim$2. This reflects the modest shielding by the small measured C mass. For larger C masses, the difference can be more dramatic.

\subsection{Implications for Debris Disks}\label{subsec:implications}
When was the dusty disk around HD 32297 produced? Is it as old as the star, or is it the result of a recent catastrophic event? What is the total mass of the underlying planetesimal belt? Our model without C removal requires a catastrophic event 1000\,yr ago, i.e.\ the CO gas was released within an orbital timescale. In such a scenario, the symmetric appearance of the disk in CO and \ion{C}{1} is challenging to explain. On the other hand, our steady state model cannot give a unique answer to the event time, so we look to other evidence for guidance.

If the CO gas has been steadily produced over a long timescale, it and its descendants (carbon and oxygen gas) ought to have viscously spread out and eventually accreted onto the star. For a gas belt of width $\Delta r$, and an $\alpha$ parameter of $0.01$, as likely appropriate for gaseous debris disks \citep{Kral_Latter_2016}, the time required to double the radial width is of order
\begin{equation}\label{eq:tspread}
    t \sim \frac{(\Delta r)^2}{\alpha c_s H} \sim 10^6\,{\rm yr}\, \left( \frac{\Delta r}{40\,{\rm au}}\right)^2.
\end{equation}
However, our modeling of the \ion{C}{1} emission (section \ref{sec:CI_modeling}) suggests that the gas is distributed in a ring that is only mildly broader than the dust ring, indicating that there is insignificant radial spreading around HD\,32297, since gas production started. Could this imply that the event time is shorter than the system lifetime ($\geq 15$ Myrs)? The answer is no. Any newly outgassed CO is necessarily concentrated near the grains' orbits. It is photo-dissociated within a time short compared to the above spreading time (mean lifetime in our model is $0.1$\,Myr). And its descendant, the C gas, is rapidly removed ($\tau_\mathrm{C}\sim700$\,yr). So no radial spreading is allowed in the steady state model.

The high CO production rate that we obtain ($\sim 15$\,M$_\oplus$\,Myr$^{-1}$) can in principle place a constraint on the event time, since it is not sustainable over a long period of time. However, this is not true if CO recycling occurs. In the latter case, even if gas production started right after the dispersal of the protoplanetary disk, the minimum CO required is still only of order M$_\oplus$ (the current CO gas mass), roughly the amount of CO contained by a comet belt of some 10\,M$_\oplus$. This total mass is consistent with the estimates based on the gradual luminosity evolution of debris disks \citep[e.g.][]{Shannon_Wu_2011}.

Previously, \citet{Cataldi_etal_2018}  concluded that the debris disk around $\beta$~Pic has to be produced recently. In that system, the $^{12}$CO emission is optically thin and the CO mass can be directly measured \citep{Matra_etal_2017_betaPic}. It is at least three orders of magnitude lower than that in HD\,32297, while the carbon masses are comparable. Moreover, both CO and carbon are distributed in a highly asymmetric, radially confined ring \citep{Cataldi_etal_2018}. The low C and CO masses imply little shielding and the CO lifetime is as short as $\sim$50\,yr, or a CO destruction rate of  $0.6$\,M$_\oplus$\,Myr$^{-1}$. Assuming that the low amount of C is the integrated result of this destruction rate, \citet{Cataldi_etal_2018} concluded that carbon can be accumulated in a time as short as $\lesssim10^4$\,yr. However, this picture of a short event time would drastically be changed if C is removed. In such a case, the low carbon mass no longer places a constraint on the event time directly.

Besides $\beta$~Pic and HD\,32297, there are currently two other debris disks with resolved CO and \ion{C}{1} observations: HD~131835 and 49~Ceti. For HD\,131835, \citet{Kral_etal_2019} derived $r_0=90$\,au, $\Delta r=70$\,au and a C$^0$ mass between $2.7\times10^{-3}$\,M$_\oplus$ and $1.2\times10^{-2}$\,M$_\oplus$. The CO mass is estimated at $0.04$\,M$_\oplus$ \citep{Moor_etal_2017,Kral_etal_2019}. The spatial distribution of the gas is uncertain and there remains the possibility that an accretion disk has formed \citep{Kral_etal_2019}. For 49~Ceti, the dust belt is concentrated around $r_0 \sim 100$AU, with a width $\Delta r\sim80$\,au \citep[the double power law model of][]{Hughes_etal_2017}, while the carbon mass is $M_\mathrm{C^0}=3\times10^{-3}$\,M$_\oplus$ (assuming optically thin emission; A.\ Higuchi, private communication 2019) and the CO mass (if ISM ratios apply) $(1.11\pm0.13)\times10^{-2}$\,M$_\oplus$ \citep{Moor_etal_2019}. \citet{Hughes_etal_2017} found that the CO observations are well described by a model where the surface density increases with radius between $\sim$20 and $\sim$220\,au, suggesting that an accretion disk is unlikely. Both disks might be qualitatively explained by a scenario similar to HD~32297 (figure \ref{fig:gas_evolution_fitting_models} right).

Recently, \citet{Marino_etal_2020} presented a population synthesis study by coupling the secondary gas production model by \citet{Kral_etal_2019} with a dust collisional evolution model (i.e. the dust grind down rate, and thus CO production, decay with time). While they find agreement with measured CO masses, their model tends to overpredict the masses of neutral C, although they get better agreement by assuming that C beyond the belt location is not detectable in observations. This result seems analogous to our finding that the small C mass around HD~32297 is difficult to reproduce with a CO production rate derived from the dust mass loss rate.

\section{Summary}\label{sec:summary}
In this work, we present resolved ALMA observations of \ion{C}{1} emission at 492\,GHz from the HD\,32297 debris disk. The data are used to investigate the age and the gas production rate of the disk. Our results can be summarized as follows:

\begin{enumerate}

    \item The spatial distribution of neutral carbon can be modelled by a Gaussian ring surface density profile centered at 110\,au with a FWHM of 80\,au. A double power law model can also fit the data satisfactorily. The mass of neutral carbon is $(3.5\pm0.2)\times10^{-3}$\,M$_\oplus$.
    
    \item We attempt to explain the gas-rich debris disk around HD~32297 by secondary gas production. We present a model where CO is produced at a fixed rate and destroyed by photodissociation, which in turn produces carbon. The carbon can itself be removed (e.g.\ by capture onto dust grains). The model includes CO self-shielding and CO shielding by neutral carbon.
    
    \item We aim to reproduce both the C$^0$ mass determined in this work and the C$^{18}$O mass. Without removal of C gas, our model requires that 0.1\,M$_\oplus$ of CO were released over the past 1000\,yr, maybe in the catastrophic destruction of a planetary body. However, the absence of asymmetries in the \ion{C}{1} and CO data might be hard to explain in such a scenario. For a steady state solution, the model requires very efficient removal of C ($\tau_\mathrm{C}\sim700$\,yr). Even in that case, the CO production rate remains high at $\sim$15\,M$_\oplus$\,Myr$^{-1}$, exceeding the dust mass removal rate. For such high CO production rates, shielding by C is negligible compared to CO self-shielding. The total CO mass inferred from our model in steady state is 1.6\,M$_\oplus$.

    \item The existence of a steady-state solution means that the event time (the time when dust and gas production commence) is not constrained in our model.
    
    \item We consider the fate of C$^{0}$ atoms that collide with grains. Depending on the binding energy, they may stick to the grains. We  propose that they might reform CO (with incoming O atoms) that is re-released to the gas phase. This would lead to a continuous recycling of CO. Such a picture would explain the high CO production rate, and the belt-like gas disk, despite viscous spreading.

\end{enumerate}

It is desirable to verify our model and to extend it to other debris disks with gas observations. In particular, the idea of CO re-formation needs to be tested, as it might be an important process in gas-rich debris disks. And there still exists a discrepancy between our inferred carbon removal timescale and that estimated based on dust grain capture.

\acknowledgments
We thank the anonymous referee for constructive comments that significantly improved the structure and clarity of our paper. We acknowledge useful discussions with Yuri Aikawa, Per Calissendorff, Misato Fukagawa, Kenji Furuya, Ryohei Kawabe, Luca Matr\`a, Shoji Mori, Nami Sakai, Alessandro Trani, Takashi Tsukagoshi and Satoshi Yamamoto. We thank Alan Heays for providing photodissociation cross sections of CO. Gianni Cataldi is supported by NAOJ ALMA Scientific Research Grant Number 2019-13B. Attila Mo{\'o}r acknowledges the support of the Hungarian National Research, Development and Innovation Office NKFIH Grant KH-130526. Thomas Henning acknowledges support from the European Research Council under the Horizon 2020 Framework Program via the ERC Advanced Grant Origins 83 24 28. This project has received funding from the European Research Council (ERC) under the European Union's Horizon 2020 research and innovation programme under grant agreement No 716155 (SACCRED). This paper makes use of the following ALMA data: ADS/JAO.ALMA\#2017.1.00201.S. ALMA is a partnership of ESO (representing its member states), NSF (USA) and NINS (Japan), together with NRC (Canada), MOST and ASIAA (Taiwan), and KASI (Republic of Korea), in cooperation with the Republic of Chile. The Joint ALMA Observatory is operated by ESO, AUI/NRAO and NAOJ. This research has made use of NASA’s Astrophysics Data System, and the SIMBAD database operated at CDS, Strasbourg, France.

%

\vspace{5mm}
\facilities{ALMA}


\software{astropy \citep{astropy_2018}, CASA \citep{McMullin_etal_2007}, corner.py \citep{Foreman-Mackey_2016}, emcee \citep{Foreman-Mackey_etal_2013}, LIME \citep{Brinch_Hogerheijde_2010}, Matplotlib \citep{Hunter_2007}, NumPy \citep{vanderWalt_etal_2011}, pythonradex (\url{https://github.com/gica3618/pythonradex}), SciPy \citep{Virtanen_etal_2020}}



\appendix
\section{Position angle of the disk}\label{appendix:PA}
The position angle (PA) of the HD\,32297 disk was measured from ALMA Band 6 continuum data by \citet{MacGregor_etal_2018}. They found $47.9\arcdeg\pm0.2\arcdeg$, consistent with measurements from scattered light observations \citep[e.g.][]{Asensio-Torres_etal_2016}. We measure the PA from our continuum image (Fig.~\ref{fig:CI_mom0_continuum}, right). For a given trial PA $\phi$, we first rotate the disk by an angle $\pi-\phi$ to align it with the horizontal axis. Then, we use three different methods to grade the trial PA:
\begin{enumerate}
\item We consider the flux inside a $2\arcsec\times0.4\arcsec$ box (this is slightly smaller than the extent of the disk) centered at the stellar position. We search for the PA that maximizes the flux.
\item We follow \citet{Asensio-Torres_etal_2016} and mirror the rotated image along the vertical axis. Then, the mirror image is subtracted from the rotated image to get a residual image. We search for the PA $\phi$ that minimizes the residuals.
\item We follow \citet{Matra_etal_2017_betaPic} and fit a Gaussian to each vertical row of pixels in the rotated image. The centers of the Gaussians define the spine of the disk. We perform a linear fit to the spine. We search for the PA that produces a linear fit to the spine with a slope closest to zero.
\end{enumerate}
We consider trial PAs between $43\arcdeg$ and $53\arcdeg$ with steps of $0.2\arcdeg$, giving us an estimate of the PA for each of the three methods. In order to estimate errors, we employ a Monte Carlo method and resample the data 1000 times by adding noise to our image and repeating the fit \citep[e.g.][]{Andrae_2010}. The noise in our image is correlated. We extend the method described in Appendix~A of \citet{Cataldi_etal_2014} to two dimensions to produce synthetic correlated noise that can be added to our image. We find PA values of $49.2\arcdeg\pm1.0\arcdeg$, $48.2\arcdeg\pm1.3\arcdeg$ and $48.2\arcdeg\pm1.0\arcdeg$ for each of the above methods respectively. Taking the mean and combining the errors quadratically yields a final estimate of $48.5\arcdeg\pm0.6\arcdeg$, consistent with the value derived by \citet{MacGregor_etal_2018}. We repeat the analysis for the \ion{C}{1} moment 0 map (Fig.~\ref{fig:CI_mom0_continuum}, left) and find values consistent with the PA estimated from the continuum.

\section{Power law models}\label{appendix:power_law_models}
\begin{figure}
\plotone{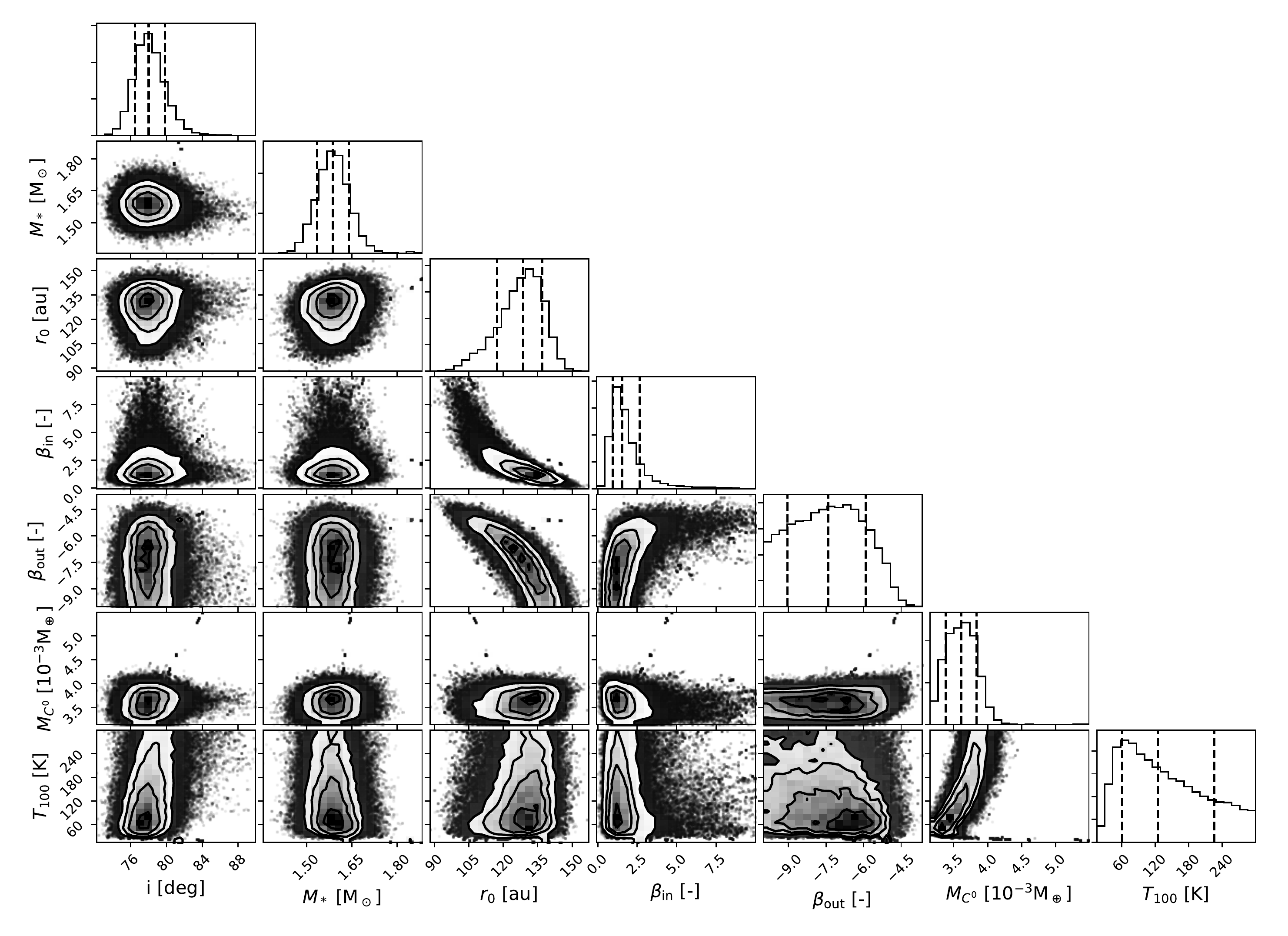}
\caption{Posterior distributions for selected parameters of the double power law model. The vertical dashed lines indicate the 16th, 50th and 84th percentile.\label{fig:cornerplot_DPL}}
\end{figure}

\begin{figure}
\plotone{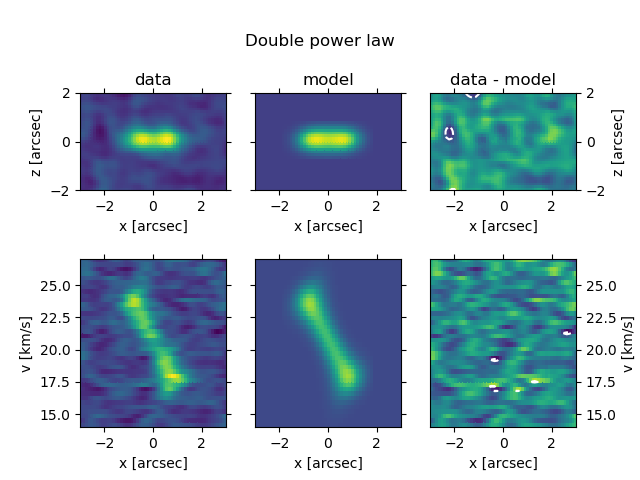}
\caption{Same as Fig.~\ref{fig:highest_prop_Gaussian_ring}, but for the double power law model. The parameter values for this model are as follows: $i=77.2$\arcdeg, $r_0=136$\,au, $\beta_\mathrm{in}=1.1$, $\beta_\mathrm{out}=-9.0$, $T_{100}=69$\,K, $M_\mathrm{C^0}=3.3\times10^{-3}$\,M$_\oplus$, $M_*=1.6$\,M$_\odot$, $v_*=20.7$\,km\,s$^{-1}$, $\Delta x=-0.05$\arcsec, $\Delta y=0.04$\arcsec.\label{fig:highest_prop_DPL}}
\end{figure}

We also fit the double power law proposed by \citet{Kral_etal_2019} to our data. The surface density of neutral carbon is given by
\begin{equation}
    \Sigma_\mathrm{C^0}(r) = 
    \begin{cases}
    \Sigma_0\left(\frac{r}{r_0}\right)^{\beta_\mathrm{in}},& \text{if } r<r_0\\
    \Sigma_0\left(\frac{r}{r_0}\right)^{\beta_\mathrm{out}},& \text{if } r>r_0.
    \end{cases}
\end{equation}
The fitted parameters remain the same as for the Gaussian ring, except that $\sigma_r$ is replaced by $\beta_\mathrm{in}$ and $\beta_\mathrm{out}$. Table~\ref{tab:fit_parameters} lists the derived parameter values. Figures \ref{fig:cornerplot_DPL} and \ref{fig:highest_prop_DPL} show the posterior distribution of selected parameters and the fit with the highest posterior probability, respectively. From the posterior probability distribution we find that the parameter $\beta_\mathrm{in}>1$ at the 81\% confidence level, suggesting that the inner region is devoid of gas, as seen for the Gaussian ring fit. The parameter $\beta_\mathrm{out}$ is strongly negative, implying a sharp cutoff of the gas density at $r_0$.

In order to further test the inner extend of the disk, we fit a third model that has the same surface density as the double power law, except that no gas is present inside of a minimum radius $r_\mathrm{min}$. Also, since $r_\mathrm{min}$ and $\beta_\mathrm{in}>0$ have a similar effect, we force $\beta_\mathrm{in}$ to be negative. We find that $r_\mathrm{min}>60$\,au at the 95\% confidence level, giving further confidence that the gas is distributed in a ring-like geometry.

\section{Comparison to \ion{C}{2} data}\label{appendix:CII_comparison}
\citet{Donaldson_etal_2013} detected \ion{C}{2} 158\,$\mu$m emission towards HD\,32297. We check whether our C$^0$ models are consistent with this measurement. For a given model describing the distribution of neutral carbon, we calculate the ionisation in every grid cell of the model. For simplicity, we assume that the gas is optically thin to ionising radiation. This assumption is justified, because the ISRF is the dominating source of ionising photons and the typical vertical optical depth of our models to ionising radiation is only $\sim$1. Note however that, although neglected here, the optical depth of C$^0$ to ionising radiation is important when considering the shielding of CO (section \ref{sec:time_evolution_model}). Assuming that electrons are solely produced by C ionisation, the C$^+$ number density is given by
\begin{equation}\label{eq:ionisation}
    n_\mathrm{C^+} = \sqrt{\frac{n_\mathrm{C^0}\Gamma}{\xi}}
\end{equation}
where $\Gamma$ is the ionisation rate computed using cross sections from the NORAD database\footnote{\url{http://www.astronomy.ohio-state.edu/~nahar/nahar\_radiativeatomicdata/index.html}} \citep{Nahar_Pradhan_1991} and $\xi$ the recombination rate, taken from the same database \citep{Nahar_1995,Nahar_1996}. Assuming again LTE and raytracing the \ion{C}{2} as described in section \ref{sec:CI_modeling}, we can compute the total \ion{C}{2} flux. We find \ion{C}{2} fluxes of $2.74\times10^{-18}$\,W\,m$^{-1}$ and $2.44\times10^{-18}$\,W\,m$^{-1}$ for the Gaussian ring model of Fig.~\ref{fig:highest_prop_Gaussian_ring} and the double power law model of Fig.~\ref{fig:highest_prop_DPL} respectively. This compares well to the \citet{Donaldson_etal_2013} measurement of $2.68\pm0.72\times10^{-18}$\,W\,m$^{-2}$. The typical ionisation fraction in the mid-plane at $r_0$ of these models is $\sim$0.2.

We point out that the lower limit on the C$^+$ column density derived by \citet{Donaldson_etal_2013} is too low by a factor $\sim$10$^6$, presumably because of a computation error. Also, their equation 2 is incorrect: the factor $g_1$ should be removed.

Is it possible that an accretion disk is present, but that carbon is largely ionised in the inner regions and thus eludes detection by ALMA? Given that the ISRF dominates the ionisation of C except in close vicinity to the star ($r\lesssim 12$\,au), the inner region is expected to actually be more neutral in an accretion disk, thanks to the higher recombination rate in that denser part.

\section{Uncertainty of C mass due to low kinetic temperature}\label{appendix:mass_uncertainty_low_Tkin}
Here we investigate whether the mass of neutral carbon could be higher than determined in section \ref{sec:CI_modeling} due to low kinetic temperature. We ran another MCMC for the Gaussian ring model where we forced $T_{100}<15$\,K and fixed the parameters $i$, $M_*$, $v_*$, $\Delta x$ and $\Delta z$ to their 50th percentile value as listed in Table~\ref{tab:fit_parameters}. The 99th percentile of the C$^0$ mass distribution derived from this additional MCMC is 0.01\,M$_\oplus$, only a factor $\sim$3 larger than the 50th percentile from our original MCMC.

In addition, an argument against a low temperature comes from the observed \ion{C}{2} flux \citep{Donaldson_etal_2013}. At low temperature, the \ion{C}{2} 158\,$\mu$m line becomes hard to excite. We can derive a lower limit on the kinetic temperature by considering the maximum \ion{C}{2} flux that can possibly be emitted for a given kinetic temperature $T_\mathrm{kin}$:
\begin{equation}\label{eq:max_CII_flux}
F_\mathrm{max}=B_\nu(T_\mathrm{kin})\Omega\Delta\nu
\end{equation}
where $B_\nu$ is the Planck function, $\Omega$ the solid angle of the emitting region and $\Delta\nu=\Delta v\nu_0/c$ the width of the emission line (we assume $\Delta v=9$\,km\,s$^{-1}$, see Fig.~\ref{fig:pv_diagram}). Equation \ref{eq:max_CII_flux} corresponds to completely optically thick emission. Assuming $\Omega=2r_\mathrm{max}4H/d^2$ (i.e.\ an edge-on disk extending out to $r_\mathrm{max}$ in the radial direction and two scale heights $H$ in the vertical direction, seen at distance $d$) with $r_\mathrm{max}=150$\,au and $H=7$\,au, we find that the \ion{C}{2} flux observed by \citet{Donaldson_etal_2013} can only be reproduced if $T_\mathrm{kin}>28$\,K.

Of course, if there is a strong temperature gradient in the disk, mass could still be hiding in low temperature regions.

\section{Alternative derivation of the CO destruction rate}\label{appendix:alternative_CO_destruction_rate_derivation}
Here we present an alternative derivation of the CO destruction rate. Let a CO molecule be located at a depth $z$. The probability that a given photon will survive to the depth $z$ is $\exp(-\tau_\mathrm{tot}(z))$. Thus, the probability that this ISRF photon will destroy this particular CO molecule, is
\begin{equation}
    p_i = \exp(-\tau_\mathrm{tot}(z))\frac{\sigma_\mathrm{CO}}{A}.
\end{equation}
To get the probability that the photon destroys any of the CO molecules in the disk, we have to sum over all CO molecules:
\begin{equation}
    p(\mathrm{CO})=\sum_i p_i = \sigma_\mathrm{CO}n_\mathrm{CO}\int\exp(-\tau_\mathrm{tot}(z))\mathrm{d}z
\end{equation}
Using $\tau_\mathrm{tot}(z)=z(n_\mathrm{C^0}\sigma_\mathrm{C^0}+n_\mathrm{CO}\sigma_\mathrm{CO})$, this expression is easily evaluated to be $(1-\exp(-\tau_\mathrm{tot}))\tau_\mathrm{CO}/\tau_\mathrm{tot}$, from which equation \ref{eq:CO_destruction_rate}, which integrates over all available photons, follows.

\bibliographystyle{aasjournal}
\bibliography{bibliography}

\begin{thebibliography}{}
\expandafter\ifx\csname natexlab\endcsname\relax\def\natexlab#1{#1}\fi
\providecommand{\url}[1]{\href{#1}{#1}}

\bibitem[{{Altwegg} {et~al.}(2019){Altwegg}, {Balsiger}, \&
  {Fuselier}}]{Altwegg_etal_2019}
{Altwegg}, K., {Balsiger}, H., \& {Fuselier}, S.~A. 2019, \araa, 57, 113

\bibitem[{{Andrae}(2010)}]{Andrae_2010}
{Andrae}, R. 2010, ArXiv e-prints, arXiv:1009.2755

\bibitem[{{Armitage}(2009)}]{Armitage_2009}
{Armitage}, P.~J. 2009, {Astrophysics of Planet Formation} (Cambridge
  University Press)

\bibitem[{{Asensio-Torres} {et~al.}(2016){Asensio-Torres}, {Janson},
  {Hashimoto}, {Thalmann}, {Currie}, {Buenzli}, {Kudo}, {Kuzuhara}, {Kusakabe},
  {Abe}, {Akiyama}, {Brandner}, {Brandt}, {Carson}, {Egner}, {Feldt}, {Goto},
  {Grady}, {Guyon}, {Hayano}, {Hayashi}, {Hayashi}, {Henning}, {Hodapp},
  {Ishii}, {Iye}, {Kandori}, {Knapp}, {Kwon}, {Matsuo}, {McElwain}, {Mayama},
  {Miyama}, {Morino}, {Moro-Martin}, {Nishimura}, {Pyo}, {Serabyn}, {Suenaga},
  {Suto}, {Suzuki}, {Takahashi}, {Takami}, {Takato}, {Terada}, {Turner},
  {Watanabe}, {Wisniewski}, {Yamada}, {Takami}, {Usuda}, \&
  {Tamura}}]{Asensio-Torres_etal_2016}
{Asensio-Torres}, R., {Janson}, M., {Hashimoto}, J., {et~al.} 2016, \aap, 593,
  A73

\bibitem[{{Astropy Collaboration} {et~al.}(2018){Astropy Collaboration},
  {Price-Whelan}, {Sip{\H{o}}cz}, {G{\"u}nther}, {Lim}, {Crawford}, {Conseil},
  {Shupe}, {Craig}, {Dencheva}, {Ginsburg}, {Vand erPlas}, {Bradley},
  {P{\'e}rez-Su{\'a}rez}, {de Val-Borro}, {Aldcroft}, {Cruz}, {Robitaille},
  {Tollerud}, {Ardelean}, {Babej}, {Bach}, {Bachetti}, {Bakanov}, {Bamford},
  {Barentsen}, {Barmby}, {Baumbach}, {Berry}, {Biscani}, {Boquien}, {Bostroem},
  {Bouma}, {Brammer}, {Bray}, {Breytenbach}, {Buddelmeijer}, {Burke},
  {Calderone}, {Cano Rodr{\'\i}guez}, {Cara}, {Cardoso}, {Cheedella}, {Copin},
  {Corrales}, {Crichton}, {D'Avella}, {Deil}, {Depagne}, {Dietrich}, {Donath},
  {Droettboom}, {Earl}, {Erben}, {Fabbro}, {Ferreira}, {Finethy}, {Fox},
  {Garrison}, {Gibbons}, {Goldstein}, {Gommers}, {Greco}, {Greenfield},
  {Groener}, {Grollier}, {Hagen}, {Hirst}, {Homeier}, {Horton}, {Hosseinzadeh},
  {Hu}, {Hunkeler}, {Ivezi{\'c}}, {Jain}, {Jenness}, {Kanarek}, {Kendrew},
  {Kern}, {Kerzendorf}, {Khvalko}, {King}, {Kirkby}, {Kulkarni}, {Kumar},
  {Lee}, {Lenz}, {Littlefair}, {Ma}, {Macleod}, {Mastropietro}, {McCully},
  {Montagnac}, {Morris}, {Mueller}, {Mumford}, {Muna}, {Murphy}, {Nelson},
  {Nguyen}, {Ninan}, {N{\"o}the}, {Ogaz}, {Oh}, {Parejko}, {Parley}, {Pascual},
  {Patil}, {Patil}, {Plunkett}, {Prochaska}, {Rastogi}, {Reddy Janga},
  {Sabater}, {Sakurikar}, {Seifert}, {Sherbert}, {Sherwood-Taylor}, {Shih},
  {Sick}, {Silbiger}, {Singanamalla}, {Singer}, {Sladen}, {Sooley},
  {Sornarajah}, {Streicher}, {Teuben}, {Thomas}, {Tremblay}, {Turner},
  {Terr{\'o}n}, {van Kerkwijk}, {de la Vega}, {Watkins}, {Weaver}, {Whitmore},
  {Woillez}, {Zabalza}, \& {Astropy Contributors}}]{astropy_2018}
{Astropy Collaboration}, {Price-Whelan}, A.~M., {Sip{\H{o}}cz}, B.~M., {et~al.}
  2018, \aj, 156, 123

\bibitem[{{Bailer-Jones} {et~al.}(2018){Bailer-Jones}, {Rybizki}, {Fouesneau},
  {Mantelet}, \& {Andrae}}]{BailerJones_etal_2018}
{Bailer-Jones}, C.~A.~L., {Rybizki}, J., {Fouesneau}, M., {Mantelet}, G., \&
  {Andrae}, R. 2018, \aj, 156, 58

\bibitem[{{Bhowmik} {et~al.}(2019){Bhowmik}, {Boccaletti}, {Th{\'e}bault},
  {Kral}, {Mazoyer}, {Milli}, {Maire}, {van Holstein}, {Augereau}, {Baudoz},
  {Feldt}, {Galicher}, {Henning}, {Lagrange}, {Olofsson}, {Pantin}, \&
  {Perrot}}]{Bhowmik_etal_2019}
{Bhowmik}, T., {Boccaletti}, A., {Th{\'e}bault}, P., {et~al.} 2019, \aap, 630,
  A85

\bibitem[{{Boccaletti} {et~al.}(2012){Boccaletti}, {Augereau}, {Lagrange},
  {Milli}, {Baudoz}, {Mawet}, {Mouillet}, {Lebreton}, \&
  {Maire}}]{Boccaletti_etal_2012}
{Boccaletti}, A., {Augereau}, J.~C., {Lagrange}, A.~M., {et~al.} 2012, \aap,
  544, A85

\bibitem[{{Booth} {et~al.}(2017){Booth}, {Dent}, {Jord{\'a}n}, {Lestrade},
  {Hales}, {Wyatt}, {Casassus}, {Ertel}, {Greaves}, {Kennedy}, {Matr{\`a}},
  {Augereau}, \& {Villard}}]{Booth_etal_2017}
{Booth}, M., {Dent}, W. R.~F., {Jord{\'a}n}, A., {et~al.} 2017, \mnras, 469,
  3200

\bibitem[{{Bouret} {et~al.}(2002){Bouret}, {Deleuil}, {Lanz}, {Roberge},
  {Lecavelier des Etangs}, \& {Vidal-Madjar}}]{Bouret_etal_2002}
{Bouret}, J.~C., {Deleuil}, M., {Lanz}, T., {et~al.} 2002, \aap, 390, 1049

\bibitem[{{Brinch} \& {Hogerheijde}(2010)}]{Brinch_Hogerheijde_2010}
{Brinch}, C., \& {Hogerheijde}, M.~R. 2010, \aap, 523, A25

\bibitem[{{Cataldi} {et~al.}(2014){Cataldi}, {Brandeker}, {Olofsson},
  {Larsson}, {Liseau}, {Blommaert}, {Fridlund}, {Ivison}, {Pantin},
  {Sibthorpe}, {Vandenbussche}, \& {Wu}}]{Cataldi_etal_2014}
{Cataldi}, G., {Brandeker}, A., {Olofsson}, G., {et~al.} 2014, \aap, 563, A66

\bibitem[{{Cataldi} {et~al.}(2018){Cataldi}, {Brandeker}, {Wu}, {Chen}, {Dent},
  {de Vries}, {Kamp}, {Liseau}, {Olofsson}, {Pantin}, \&
  {Roberge}}]{Cataldi_etal_2018}
{Cataldi}, G., {Brandeker}, A., {Wu}, Y., {et~al.} 2018, \apj, 861, 72

\bibitem[{{Currie} {et~al.}(2012){Currie}, {Rodigas}, {Debes}, {Plavchan},
  {Kuchner}, {Jang-Condell}, {Wilner}, {Andrews}, {Kraus}, {Dahm}, \&
  {Robitaille}}]{Currie_etal_2012}
{Currie}, T., {Rodigas}, T.~J., {Debes}, J., {et~al.} 2012, \apj, 757, 28

\bibitem[{{Deleuil} {et~al.}(2001){Deleuil}, {Bouret}, {Lecavelier des Etangs},
  {Roberge}, {Vidal-Madjar}, {Andr{\'e}}, {Blair}, {Feldman}, {Ferlet},
  {Friedman}, \& {Moos}}]{Deleuil_etal_2001}
{Deleuil}, M., {Bouret}, J.~C., {Lecavelier des Etangs}, A., {et~al.} 2001,
  \apj, 557, L67

\bibitem[{{Donaldson} {et~al.}(2013){Donaldson}, {Lebreton}, {Roberge},
  {Augereau}, \& {Krivov}}]{Donaldson_etal_2013}
{Donaldson}, J.~K., {Lebreton}, J., {Roberge}, A., {Augereau}, J.~C., \&
  {Krivov}, A.~V. 2013, \apj, 772, 17

\bibitem[{{Draine}(1978)}]{Draine_1978}
{Draine}, B.~T. 1978, The Astrophysical Journal Supplement Series, 36, 595

\bibitem[{{Fitzgerald} {et~al.}(2007){Fitzgerald}, {Kalas}, \&
  {Graham}}]{Fitzgerald_etal_2007}
{Fitzgerald}, M.~P., {Kalas}, P.~G., \& {Graham}, J.~R. 2007, \apj, 670, 557

\bibitem[{{Fomalont} {et~al.}(2014){Fomalont}, {van Kempen}, {Kneissl},
  {Marcelino}, {Barkats}, {Corder}, {Cortes}, {Hills}, {Lucas}, \&
  {Manning}}]{Fomalont_etal_2014}
{Fomalont}, E., {van Kempen}, T., {Kneissl}, R., {et~al.} 2014, The Messenger,
  155, 19

\bibitem[{{Foreman-Mackey}(2016)}]{Foreman-Mackey_2016}
{Foreman-Mackey}, D. 2016, The Journal of Open Source Software, 1, 24

\bibitem[{{Foreman-Mackey} {et~al.}(2013){Foreman-Mackey}, {Hogg}, {Lang}, \&
  {Goodman}}]{Foreman-Mackey_etal_2013}
{Foreman-Mackey}, D., {Hogg}, D.~W., {Lang}, D., \& {Goodman}, J. 2013,
  Publications of the Astronomical Society of the Pacific, 125, 306

\bibitem[{{Glover} {et~al.}(2010){Glover}, {Federrath}, {Mac Low}, \&
  {Klessen}}]{Glover_etal_2010}
{Glover}, S.~C.~O., {Federrath}, C., {Mac Low}, M.-M., \& {Klessen}, R.~S.
  2010, \mnras, 404, 2

\bibitem[{{Greaves} {et~al.}(2016){Greaves}, {Holland}, {Matthews}, {Marshall},
  {Dent}, {Woitke}, {Wyatt}, {Matr{\`a}}, \& {Jackson}}]{Greaves_etal_2016}
{Greaves}, J.~S., {Holland}, W.~S., {Matthews}, B.~C., {et~al.} 2016, \mnras,
  461, 3910

\bibitem[{{Grigorieva} {et~al.}(2007){Grigorieva}, {Th{\'e}bault},
  {Artymowicz}, \& {Brandeker}}]{Grigorieva_etal_2007}
{Grigorieva}, A., {Th{\'e}bault}, P., {Artymowicz}, P., \& {Brandeker}, A.
  2007, \aap, 475, 755

\bibitem[{{H{\"a}ssig} {et~al.}(2017){H{\"a}ssig}, {Altwegg}, {Balsiger},
  {Berthelier}, {Bieler}, {Calmonte}, {Dhooghe}, {Fiethe}, {Fuselier}, {Gasc},
  {Gombosi}, {Le Roy}, {Luspay-Kuti}, {Mand t}, {Rubin}, {Tzou}, {Wampfler}, \&
  {Wurz}}]{Hassig_etal_2017}
{H{\"a}ssig}, M., {Altwegg}, K., {Balsiger}, H., {et~al.} 2017, \aap, 605, A50

\bibitem[{{Heays} {et~al.}(2017){Heays}, {Bosman}, \& {van
  Dishoeck}}]{Heays_etal_2017}
{Heays}, A.~N., {Bosman}, A.~D., \& {van Dishoeck}, E.~F. 2017, \aap, 602, A105

\bibitem[{{Henry}(2002)}]{Henry_2002}
{Henry}, R.~C. 2002, \apj, 570, 697

\bibitem[{{Hollenbach} \& {Salpeter}(1970)}]{Hollenbach_Salpeter_1970}
{Hollenbach}, D., \& {Salpeter}, E.~E. 1970, \jcp, 53, 79

\bibitem[{{Huebner}(2002)}]{Huebner_2002}
{Huebner}, W.~F. 2002, Earth Moon and Planets, 89, 179

\bibitem[{{Hughes} {et~al.}(2018){Hughes}, {Duch{\^e}ne}, \&
  {Matthews}}]{Hughes_etal_2018}
{Hughes}, A.~M., {Duch{\^e}ne}, G., \& {Matthews}, B.~C. 2018, Annual Review of
  Astronomy and Astrophysics, 56, 541

\bibitem[{{Hughes} {et~al.}(2017){Hughes}, {Lieman-Sifry}, {Flaherty}, {Daley},
  {Roberge}, {K{\'o}sp{\'a}l}, {Mo{\'o}r}, {Kamp}, {Wilner}, {Andrews},
  {Kastner}, \& {{\'A}brah{\'a}m}}]{Hughes_etal_2017}
{Hughes}, A.~M., {Lieman-Sifry}, J., {Flaherty}, K.~M., {et~al.} 2017, \apj,
  839, 86

\bibitem[{Hunter(2007)}]{Hunter_2007}
Hunter, J.~D. 2007, Computing In Science \& Engineering, 9, 90

\bibitem[{Jaworski {et~al.}(2016)Jaworski, Zakrzewska, \&
  Pianko-Oprych}]{Jaworski_etal_2016}
Jaworski, Z., Zakrzewska, B., \& Pianko-Oprych, P. 2016, Reviews in Chemical
  Engineering, 33, doi:10.1515/revce-2016-0022

\bibitem[{{Kalas}(2005)}]{Kalas_etal_2005}
{Kalas}, P. 2005, \apj, 635, L169

\bibitem[{{K{\'o}sp{\'a}l} {et~al.}(2013){K{\'o}sp{\'a}l}, {Mo{\'o}r},
  {Juh{\'a}sz}, {{\'A}brah{\'a}m}, {Apai}, {Csengeri}, {Grady}, {Henning},
  {Hughes}, {Kiss}, {Pascucci}, \& {Schmalzl}}]{Kospal_etal_2013}
{K{\'o}sp{\'a}l}, {\'A}., {Mo{\'o}r}, A., {Juh{\'a}sz}, A., {et~al.} 2013,
  \apj, 776, 77

\bibitem[{{Kral} \& {Latter}(2016)}]{Kral_Latter_2016}
{Kral}, Q., \& {Latter}, H. 2016, \mnras, 461, 1614

\bibitem[{{Kral} {et~al.}(2019){Kral}, {Marino}, {Wyatt}, {Kama}, \&
  {Matr{\`a}}}]{Kral_etal_2019}
{Kral}, Q., {Marino}, S., {Wyatt}, M.~C., {Kama}, M., \& {Matr{\`a}}, L. 2019,
  \mnras, 489, 3670

\bibitem[{{Kral} {et~al.}(2017){Kral}, {Matr{\`a}}, {Wyatt}, \&
  {Kennedy}}]{Kral_etal_2017}
{Kral}, Q., {Matr{\`a}}, L., {Wyatt}, M.~C., \& {Kennedy}, G.~M. 2017, \mnras,
  469, 521

\bibitem[{{Leitch-Devlin} \& {Williams}(1985)}]{Leitch-Devlin_Williams_1985}
{Leitch-Devlin}, M.~A., \& {Williams}, D.~A. 1985, \mnras, 213, 295

\bibitem[{{Lieman-Sifry} {et~al.}(2016){Lieman-Sifry}, {Hughes}, {Carpenter},
  {Gorti}, {Hales}, \& {Flaherty}}]{Lieman-Sifry_etal_2016}
{Lieman-Sifry}, J., {Hughes}, A.~M., {Carpenter}, J.~M., {et~al.} 2016, \apj,
  828, 25

\bibitem[{{MacGregor} {et~al.}(2016){MacGregor}, {Wilner}, {Chandler}, {Ricci},
  {Maddison}, {Cranmer}, {Andrews}, {Hughes}, \&
  {Steele}}]{MacGregor_etal_2016}
{MacGregor}, M.~A., {Wilner}, D.~J., {Chandler}, C., {et~al.} 2016, \apj, 823,
  79

\bibitem[{{MacGregor} {et~al.}(2018){MacGregor}, {Weinberger}, {Hughes},
  {Wilner}, {Currie}, {Debes}, {Donaldson}, {Redfield}, {Roberge}, \&
  {Schneider}}]{MacGregor_etal_2018}
{MacGregor}, M.~A., {Weinberger}, A.~J., {Hughes}, A.~M., {et~al.} 2018, \apj,
  869, 75

\bibitem[{{Maness} {et~al.}(2008){Maness}, {Fitzgerald}, {Paladini}, {Kalas},
  {Duchene}, \& {Graham}}]{Maness_etal_2008}
{Maness}, H.~L., {Fitzgerald}, M.~P., {Paladini}, R., {et~al.} 2008, \apj, 686,
  L25

\bibitem[{{Marino} {et~al.}(2020){Marino}, {Flock}, {Henning}, {Kral},
  {Matr{\`a}}, \& {Wyatt}}]{Marino_etal_2020}
{Marino}, S., {Flock}, M., {Henning}, T., {et~al.} 2020, \mnras, 492, 4409

\bibitem[{{Marino} {et~al.}(2016){Marino}, {Matr{\`a}}, {Stark}, {Wyatt},
  {Casassus}, {Kennedy}, {Rodriguez}, {Zuckerman}, {Perez}, {Dent}, {Kuchner},
  {Hughes}, {Schneider}, {Steele}, {Roberge}, {Donaldson}, \&
  {Nesvold}}]{Marino_etal_2016}
{Marino}, S., {Matr{\`a}}, L., {Stark}, C., {et~al.} 2016, \mnras, 460, 2933

\bibitem[{{Matr{\`a}} {et~al.}(2019){Matr{\`a}}, {{\"O}berg}, {Wilner},
  {Olofsson}, \& {Bayo}}]{Matra_etal_2019}
{Matr{\`a}}, L., {{\"O}berg}, K.~I., {Wilner}, D.~J., {Olofsson}, J., \&
  {Bayo}, A. 2019, \aj, 157, 117

\bibitem[{{Matr{\`a}} {et~al.}(2017{\natexlab{a}}){Matr{\`a}}, {MacGregor},
  {Kalas}, {Wyatt}, {Kennedy}, {Wilner}, {Duchene}, {Hughes}, {Pan}, {Shannon},
  {Clampin}, {Fitzgerald}, {Graham}, {Holland }, {Pani{\'c}}, \&
  {Su}}]{Matra_etal_2017_Fomalhaut}
{Matr{\`a}}, L., {MacGregor}, M.~A., {Kalas}, P., {et~al.} 2017{\natexlab{a}},
  \apj, 842, 9

\bibitem[{{Matr{\`a}} {et~al.}(2017{\natexlab{b}}){Matr{\`a}}, {Dent}, {Wyatt},
  {Kral}, {Wilner}, {Pani{\'c}}, {Hughes}, {de Gregorio-Monsalvo}, {Hales},
  {Augereau}, {Greaves}, \& {Roberge}}]{Matra_etal_2017_betaPic}
{Matr{\`a}}, L., {Dent}, W.~R.~F., {Wyatt}, M.~C., {et~al.} 2017{\natexlab{b}},
  \mnras, 464, 1415

\bibitem[{{McMullin} {et~al.}(2007){McMullin}, {Waters}, {Schiebel}, {Young},
  \& {Golap}}]{McMullin_etal_2007}
{McMullin}, J.~P., {Waters}, B., {Schiebel}, D., {Young}, W., \& {Golap}, K.
  2007, in Astronomical Society of the Pacific Conference Series, Vol. 376,
  Astronomical Data Analysis Software and Systems XVI, ed. R.~A. {Shaw},
  F.~{Hill}, \& D.~J. {Bell}, 127

\bibitem[{{Mo{\'o}r} {et~al.}(2017){Mo{\'o}r}, {Cur{\'e}}, {K{\'o}sp{\'a}l},
  {{\'A}brah{\'a}m}, {Csengeri}, {Eiroa}, {Gunawan}, {Henning}, {Hughes},
  {Juh{\'a}sz}, {Pawellek}, \& {Wyatt}}]{Moor_etal_2017}
{Mo{\'o}r}, A., {Cur{\'e}}, M., {K{\'o}sp{\'a}l}, {\'A}., {et~al.} 2017, \apj,
  849, 123

\bibitem[{{Mo{\'o}r} {et~al.}(2019){Mo{\'o}r}, {Kral}, {{\'A}brah{\'a}m},
  {K{\'o}sp{\'a}l}, {Dutrey}, {Di Folco}, {Hughes}, {Juh{\'a}sz}, {Pascucci},
  \& {Pawellek}}]{Moor_etal_2019}
{Mo{\'o}r}, A., {Kral}, Q., {{\'A}brah{\'a}m}, P., {et~al.} 2019, \apj, 884,
  108

\bibitem[{{Nahar}(1995)}]{Nahar_1995}
{Nahar}, S.~N. 1995, The Astrophysical Journal Supplement Series, 101, 423

\bibitem[{{Nahar}(1996)}]{Nahar_1996}
---. 1996, The Astrophysical Journal Supplement Series, 106, 213

\bibitem[{{Nahar} \& {Pradhan}(1991)}]{Nahar_Pradhan_1991}
{Nahar}, S.~N., \& {Pradhan}, A.~K. 1991, Physical Review A, 44, 2935

\bibitem[{{P{\'e}ricaud} {et~al.}(2017){P{\'e}ricaud}, {Di Folco}, {Dutrey},
  {Guilloteau}, \& {Pi{\'e}tu}}]{Pericaud_etal_2017}
{P{\'e}ricaud}, J., {Di Folco}, E., {Dutrey}, A., {Guilloteau}, S., \&
  {Pi{\'e}tu}, V. 2017, \aap, 600, A62

\bibitem[{{Redfield}(2007)}]{Redfield_etal_2007}
{Redfield}, S. 2007, \apj, 656, L97

\bibitem[{{Roberge} {et~al.}(2006){Roberge}, {Feldman}, {Weinberger},
  {Deleuil}, \& {Bouret}}]{Roberge_etal_2006}
{Roberge}, A., {Feldman}, P.~D., {Weinberger}, A.~J., {Deleuil}, M., \&
  {Bouret}, J.-C. 2006, \nat, 441, 724

\bibitem[{{Rodigas} {et~al.}(2014){Rodigas}, {Debes}, {Hinz}, {Mamajek},
  {Pecaut}, {Currie}, {Bailey}, {Defrere}, {De Rosa}, {Hill}, {Leisenring},
  {Schneider}, {Skemer}, {Skrutskie}, {Vaitheeswaran}, \&
  {Ward-Duong}}]{Rodigas_etal_2014}
{Rodigas}, T.~J., {Debes}, J.~H., {Hinz}, P.~M., {et~al.} 2014, \apj, 783, 21

\bibitem[{{Rollins} \& {Rawlings}(2012)}]{Rollins_Rawlings_2012}
{Rollins}, R.~P., \& {Rawlings}, J.~M.~C. 2012, \mnras, 427, 2328

\bibitem[{{Sch{\"o}ier} {et~al.}(2005){Sch{\"o}ier}, {van der Tak}, {van
  Dishoeck}, \& {Black}}]{Schoier_etal_2005}
{Sch{\"o}ier}, F.~L., {van der Tak}, F.~F.~S., {van Dishoeck}, E.~F., \&
  {Black}, J.~H. 2005, \aap, 432, 369

\bibitem[{{Shannon} \& {Wu}(2011)}]{Shannon_Wu_2011}
{Shannon}, A., \& {Wu}, Y. 2011, \apj, 739, 36

\bibitem[{{Shimonishi} {et~al.}(2018){Shimonishi}, {Nakatani}, {Furuya}, \&
  {Hama}}]{Shimonishi_etal_2018}
{Shimonishi}, T., {Nakatani}, N., {Furuya}, K., \& {Hama}, T. 2018, \apj, 855,
  27

\bibitem[{Tielens(2005)}]{Tielens_2005}
Tielens, A. G. G.~M. 2005, The Physics and Chemistry of the Interstellar Medium
  (Cambridge University Press)

\bibitem[{{Tielens} \& {Hagen}(1982)}]{Tielens_Hagen_1982}
{Tielens}, A.~G.~G.~M., \& {Hagen}, W. 1982, \aap, 114, 245

\bibitem[{{Torres} {et~al.}(2006){Torres}, {Quast}, {da Silva}, {de La Reza},
  {Melo}, \& {Sterzik}}]{Torres_etal_2006}
{Torres}, C.~A.~O., {Quast}, G.~R., {da Silva}, L., {et~al.} 2006, \aap, 460,
  695

\bibitem[{{van der Walt} {et~al.}(2011){van der Walt}, {Colbert}, \&
  {Varoquaux}}]{vanderWalt_etal_2011}
{van der Walt}, S., {Colbert}, S.~C., \& {Varoquaux}, G. 2011, Computing in
  Science and Engineering, 13, 22

\bibitem[{{van Dishoeck}(1994)}]{vanDishoeck_1994}
{van Dishoeck}, E.~F. 1994, in Astronomical Society of the Pacific Conference
  Series, Vol.~58, The First Symposium on the Infrared Cirrus and Diffuse
  Interstellar Clouds, ed. R.~M. {Cutri} \& W.~B. {Latter}, 319

\bibitem[{{Virtanen} {et~al.}(2020){Virtanen}, {Gommers}, {Oliphant},
  {Haberland}, {Reddy}, {Cournapeau}, {Burovski}, {Peterson}, {Weckesser},
  {Bright}, {van der Walt}, {Brett}, {Wilson}, {Jarrod Millman}, {Mayorov},
  {Nelson}, {Jones}, {Kern}, {Larson}, {Carey}, {Polat}, {Feng}, {Moore}, {Vand
  erPlas}, {Laxalde}, {Perktold}, {Cimrman}, {Henriksen}, {Quintero}, {Harris},
  {Archibald}, {Ribeiro}, {Pedregosa}, {van Mulbregt}, \&
  {Contributors}}]{Virtanen_etal_2020}
{Virtanen}, P., {Gommers}, R., {Oliphant}, T.~E., {et~al.} 2020, Nature
  Methods, doi:https://doi.org/10.1038/s41592-019-0686-2

\bibitem[{{Visser} {et~al.}(2009){Visser}, {van Dishoeck}, \&
  {Black}}]{Visser_etal_2009}
{Visser}, R., {van Dishoeck}, E.~F., \& {Black}, J.~H. 2009, \aap, 503, 323

\bibitem[{{Weingartner} \& {Draine}(2001)}]{Weingartner_Draine_2001}
{Weingartner}, J.~C., \& {Draine}, B.~T. 2001, \apj, 553, 581

\bibitem[{{Wilson}(1999)}]{Wilson_1999}
{Wilson}, T.~L. 1999, Reports on Progress in Physics, 62, 143

\end{thebibliography}



\end{document}